\documentstyle[psfig,11pt]{article}
\newcommand{\mnras}{\it MNRAS\rm}
\newcommand{\aj}{\it AJ\rm}
\newcommand{\apj}{\it ApJ\rm}
\newcommand{\pasp}{\it PASP\rm}
\newcommand{\pasa}{\it PASAu\rm}
\newcommand{\apjs}{\it ApJS\rm}
\newcommand{\asa}{\it A\&A\rm}
\newcommand{\asas}{\it A\&AS\rm}
\newcommand{\araa}{\it ARA\&A\rm}

\begin{document}
\pagestyle{empty}
\vspace*{0mm}
\begin{center}
\Large
\bf
Type Ia Supernovae

\vspace*{5mm}

\normalsize
Bruno Leibundgut

\vspace*{2mm}
\small
European Southern Observatory \\
Karl-Schwarzschild-Strasse 2 \\
D--85748 Garching \\
Germany \\
{\tt bleibundgut@eso.org}
\vspace*{5mm}
\normalsize


\vfill

\section*{Abstract}
\begin{minipage}[t]{140mm}
Type Ia Supernovae are in many aspects still enigmatic objects. Their
observational and theoretical exploration is in full swing, but we
still have plenty to learn about these explosions.

Recent years have already witnessed a bonanza of supernova observations. 
The increased samples from dedicated searches have allowed the
statistical investigation of Type Ia Supernovae as a class.
The observational data on Type Ia
Supernovae are very rich, but the uniform picture of a decade ago has
been replaced by several correlations which connect the maximum
luminosity with light curve shape, color evolution, spectral appearance,
and host galaxy morphology. These correlations hold across almost the
complete spectrum of Type Ia Supernovae, with a number of notable
exceptions. After 150 days past maximum, however, all observed 
objects show the same decline rate and spectrum.

The observational constraints on explosion models are still rather
sparse. Global parameters like synthesized nickel mass, total ejecta mass
and explosion energetics are within reach in the
next few years. These parameters bypass the complicated calculations
of explosion models and radiation transport. The bolometric light curves
are a handy tool to investigate the overall appearance of Type Ia
Supernovae. The nickel masses derived this way show large variations,
which combined with the dynamics from line widths, indicate that the
brighter events are also coming from more massive objects. 

The lack of accurate distances and the uncertainty in the 
correction for absorption are hampering further
progress. Improvements in these areas are vital for the detailed comparison of
luminosities and the determination of nickel masses. Coverage at 
near-infrared wavelengths for a statistical sample of Type Ia Supernovae 
will at least decrease the dependence on the absorption. Some of the most
intriguing features of Type Ia Supernovae are best observed at
these wavelengths, like the second peak in the light curve, the
depression in the {\it J} band, and the unblended [Fe~{\sc ii}] lines in the
ashes.

\vspace*{2mm}

\centerline{\noindent To appear in {\it The Astronomy and Astrophysics Review}}

\end{minipage}

\end{center}

\break
\addtocounter{page}{-1}
\pagestyle{plain}

\tableofcontents
\section{Introduction}

The interest in supernovae has risen dramatically with their application
to cosmological problems. Their unique capabilities as distance
indicators on the cosmic scale have pushed them into the limelight of
cosmology. These objects, however, are interesting and exciting in their
own right. Supernova physics relates some of the most complicated
physical processes from the explosion mechanisms to nucleosynthesis, radiation
transport, and shock physics as well as some of the most
intriguing astrophysics ranging from star formation and stellar
evolution to cosmic metal enrichment, evolution of galaxies, and the
scale and fate of the universe. 
Their brightness makes them ideal {\it stellar} tracers at large distances
and look back times. 
Solving some of the supernova specific problems will hence give us clues
on a much larger scale. 

Stellar explosions have been observed for many 
centuries. Nonetheless, supernovae are extremely rare and only six of 
them have 
been observed over the last millennium in our own Galaxy 
(Clark \& Stevenson 1977, Murdin \& Murdin
1985, van den Bergh \& Tammann 1991). 
The number of extragalactic supernova detections has grown continuously. 
Originally only searched by a few limited experiments (e.g. Zwicky
1965) with about a dozen SNe detected each year, the
topic has taken a big turn with SN~1987A and the emergence of deep
coordinated searches for distant supernovae in the last decade. There
are now nearly 200 SNe detected each year (cf. IAU Central Bureau for
Astronomical Telegrams\footnote{\tt
http://cfa-www.harvard.edu/cfa/lists/Supernovae.html}, 
Asiago Supernova Catalog\footnote{\tt
http://merlino.pd.astro.it/$\sim$supern} (Barbon et al. 1999),
Sternberg Astronomical Institute Supernova Catalog\footnote{\tt 
http://www.sai.msu.su/cgi-bin/wdb-p95/sn/sncat/form} 
(Bartunov \& Tsvetkov 1997)). 
This plethora of objects has
revealed an increasing number of peculiar, i.e. not easily classified,
supernovae. On the other hand, it has allowed us to conduct detailed
studies of specific supernova classes and find the commonalities as well
as the individuality among supernovae. The surge of data has further
provided many new constraints for the models.

It has become increasingly clear that two main classes of
supernovae (Type~Ia {\it vs.} Type~II and Type~Ib/c) with physically completely
different backgrounds exist. Originally an observational
separation according to spectral features (Minkowski 1941, 
1964, Harkness \& Wheeler 1990, Filippenko 1997a) the classification
scheme has proven to
identify physically distinct objects. On the other hand, the
classification system has not turned out to be sufficient.
Several objects defy a clear classification and have forced extensions
to the system. There is, however, a move to a more physical description
of individual objects. Especially bright, well-observed, events have
increased our understanding of the explosions.

Type Ia Supernovae (SNe~Ia) are now almost universally accepted as thermonuclear
explosions in low-mass stars (Trimble 1982, 1983, Woosley \& Weaver 1986).
All other known supernova explosions are thought to be due to 
the core collapse in massive stars.

There are many reviews on (Type Ia) Supernovae available. The most 
comprehensive books are Petschek (1990), Wheeler, Piran, \& Weinberg
(1990), Woosley (1991), 
McCray \& Wang (1996), Bludman et al. (1997), Ruiz-Lapuente et al. (1997), 
Niemeyer \& Truran (1999), and
Livio et al. (2000). Excellent reviews were given by Trimble 
(1982, 1983) and Woosley \& Weaver (1986). More recent monographs on Type
Ia in general are Wheeler et al. (1995) and  Filippenko (1997b).
Possible progenitor systems (Branch et al. 1995, Renzini 1996, Livio 1999), 
supernova classifications (Filippenko 1997a), supernova rates (van den
Bergh \& Tammann 1991), the Hubble constant from
SNe~Ia (Branch 1998) and the status of explosion models
(Hillebrandt \& Niemeyer 2000) are covered in more specific reviews.

\subsection{Classification}

Supernovae are classified by their spectrum near maximum light (see
Filippenko 1997a for a review on supernova classifications). The
Type Ia Supernovae are characterized by the complete absence of hydrogen
and helium lines and a distinct, strong absorption line near
6100\AA, which comes from a doublet of singly ionized silicon with
$\lambda\lambda$6347\AA\ and 6371\AA.
Hydrogen or helium lines never appear in the spectra of SNe~Ia at any phase 
of the evolution. Significant variations
have been observed within this scheme, but in general SNe~Ia can safely
be distinguished from any other supernovae (Filippenko 1997a).
Great care has to be taken to separate the
SN~Ia from SNe~Ib/c which can display a similar spectrum at early
phases.

Secondary classification criteria are the late-phase spectrum dominated
by forbidden iron and cobalt lines, light curve
shape, color evolution, and host galaxy morphology. 
None of these is sufficient by itself,
but may provide additional evidence for a classification. 

\subsection{Astrophysical importance}

Since SNe~Ia are possibly the main producer of iron in the universe,
they provide a clock for the metal enrichment of matter. 
The relative long progenitor life times, as compared to
massive stars which become core-collapse supernovae, provides a
convenient feature in the relative metal abundances of $\alpha-$elements
and iron-group elements (Renzini 1999).

The heating of the interstellar medium, in particular for elliptical
galaxies, depends on the SN~Ia rates and their energy input (Ciotti et
al. 1991).

As explosions of white dwarfs SNe~Ia are placed at the end of one of the
major stellar evolution channels. Although only a few white dwarfs
really explode as SNe~Ia, they still can provide important information
on the binary fraction of stars and the evolution of binary systems both
in our Galaxy (e.g. Iben \& Tutukov 1994, 1999) and as a function of
look back time (Yungelson \& Livio 1998, Ruiz-Lapuente \& Canal 1998). 

Supernovae also play an important feedback role during the early galaxy
evolution and might be responsible for substantial loss of material from
galaxies (e.g. Wyse \& Silk 1985, Burkert \& Ruiz-Lapuente 1997,
Ferrara \& Tolstoy 2000) and the regulation of the star formation process.
The contribution of SNe~Ia as opposed to core-collapse supernovae from
massive stars is, however, unclear. 

\subsection{Type Ia Supernovae and Cosmology}

Recent years have seen SNe~Ia taking center stage in observational cosmology. 
As the momentarily best distance indicator beyond the Virgo cluster, they
provide the main route to the current expansion rate (Branch 1998)
and the deceleration of the universe (Riess et al. 1998a, Perlmutter et
al. 1999). Almost all Hubble constant determinations are now involving
SNe~Ia in one form or another. Two large HST programs have adopted
SNe~Ia as their prime distance indicator beyond the reach of
Cepheid stars (Saha et al. 1999, Gibson et al. 2000, Mould et al.
2000). Although other secondary distance indicators are still discussed,
in most approaches they enter the analyses with lower weight (Mould et
al. 2000). It is also gratifying to see that a
general convergence of the value of H$_0$ to within the respective
error bars between 60 and 70 km~s$^{-1}$~Mpc$^{-1}$ has been reached.

The claim, based on SNe~Ia, for an accelerated universe (Riess et al.
1998a, Perlmutter et al. 1999) has triggered an enthralling
debate in cosmology. It will be an important next step to verify that
supernova evolution is not mimicking a signal which has
been interpreted as a cosmological constant. Other explanations of the
supernova result apart from a cosmological constant and based on
decaying particle fields have been proposed as well (for a recent review see
Kamionkowski \& Kosowsky 1999).

\section{Observational characteristics}
\label{sec:obs}

Type Ia supernovae are characterized by the absence of hydrogen and
helium and the presence of processed material, mostly calcium, silicon 
and sulphur, 
in their spectra during the peak phase. At late phases the spectrum is
dominated by emission lines of iron-group elements. 

SNe~Ia further exhibit a distinct light curve shape, 
extreme luminosity, absence of any appreciable amount of circumstellar
material, and a lack of detectable polarization.

\subsection{Observational material}

Recent years have seen a major increase in reliable SN~Ia data at all
wavelengths.
The large collection of optical data from the
Cal\'an/Tololo Supernova Survey (Hamuy et al. 1995, Hamuy et al. 1996c)
has superseded the older, inhomogeneous catalogs (Barbon et al. 1973a,
Leibundgut et al.
1991a). The Cal\'an/Tololo survey has delivered 29 SNe~Ia with at least 
one classifying spectrum and light curves in {\it BVI}. 
At the same time, many nearby supernovae have been observed with
modern methods (reference lists can be found in Filippenko 1997a, 
Branch 1998, Contardo et al. 2000). Of particular interest are also
the new observations of SN~1997br (Li et al. 1999), SN~1997cn (Turatto 
et al. 1998) and SN~1998bu
(Suntzeff et al. 1999, Jha et al. 1999, Hernandez et al. 2000).
A catalog of SN~Ia observations has been
published by Riess et al. (1999a) summarizing the data collection 
of 22 SNe~Ia in {\it BVRI}. 

Currently there are a number of supernova searches under way which
will produce more densely sampled light curves and spectroscopic
evolutions. Data are collected in a systematic way at several places.
A large program of supernova observations has been ongoing at Asiago
(Barbon et al. 1993) for many years. 
Bright supernovae are
regularly observed in {\it BVRI} and optical spectroscopy.
The group at the Center for Astrophysics is collecting data on most
bright new supernovae. In addition to the published data (Riess et al.
1999a, Jha et al. 1999) several more SNe~Ia have been observed in
{\it BVRI} and occasionally in {\it U}. For a few objects also {\it JHK} light
curves are being assembled. 
The robotic telescope at Lick Observatory (Richmond, Treffers, \&
Filippenko 1993) is now discovering new supernovae routinely. These
objects are then followed in {\it BVRI} and with spectroscopy.
The Supernova Cosmology Project (e.g. Perlmutter et al. 1997, 1999) 
has started to search for nearby
SNe~Ia to supplement the currently existing sample. In a massive
search and follow-up program involving observatories around the
globe many objects will be observed extensively
in {\it UBVRI} and with optical spectroscopy in the next few years. 
The Mount Stromlo and Sidings Springs Abell Cluster Search has found
around 50 supernovae (Reiss et al. 1998). The light curves from this
search are a combination of very wide filters (the Macho $V_M$ and $R_M$
filter set, cf. Germany et al. 1999) and regular {\it BVRI}
observations. Only a fraction of the objects has a spectroscopic
classification. So far 6 SNe~Ia have been reported (Reiss et al.
1998). 
A follow-on project to the Cal\'an/Tololo Supernova Survey has been initiated 
recently, which concentrates on finding bright SNe~Ia and follow them
in {\it UBVRIJHK} and spectroscopy. Infrared photometry and
spectroscopy has regularly been obtained at UKIRT and the IRTF (Spyromilio
et al. 1992, Meikle et al. 1996, Bowers et al. 1997, Meikle \&
Hernandez 1999, Hernandez et al. 2000). 

Many observations are also obtained by amateur astronomers. 
Collaborative efforts are undertaken by the International Supernova 
Network\footnote{\tt http://www.supernovae.net/isn.htm}, 
the Variable Star Network\footnote{\tt 
http://www.kusastro.kyoto-u.ac.jp/vsnet/SNe/SNe.html} (VSNET), and through
the Astronomy Section of the Rochester Academy of Sciences\footnote{\tt 
http://www.ggw.org/asras/snimages}. The collaboration between
professional and amateur astronomers is becoming a very valuable
extension of supernova research. The combination of observations from
all sources has played an important role in some of the recent
research projects (e.g. Riess et al. 1999b).

\subsubsection{SN~Ia Rates}

The frequency of supernovae carries important information about their
parent population and the physical process which drives the explosions.
Although the relative importance of this measurement has been recognized,
it has been very
difficult to derive good numbers for the supernova rates (for reviews
see van den Bergh \& Tammann 1991, Tammann 1994, Strom 1995). The main
problem is the extreme rarity of supernovae which makes statistically
significant samples very difficult to come by. The problem worsens when the
rates are split into many supernova and galaxy subtypes leaving very few
objects per sampling bin. 
The searches have further to consider the time for which
supernovae could have been discovered. The control time is an important
parameter which is typically difficult to calculate as it depends on
light curve shape, absolute luminosity, and the absorption by dust in
the local supernova environment (Tammann 1994, Strom 1995). 
Supernova rates are normally
expressed as the number of supernovae per century per blue unit (i.e.
solar) luminosity
(typically per $10^{10}$L$_{B_{\odot}}$). The latter reflects the belief that
supernovae are linked to the stellar population which dominates the
galaxy light. The attempt to measure the SN~Ia rate per
{\it H} luminosity, which traces older stellar populations, seems more
reasonable (van den Bergh 1990), if we believe that SNe~Ia come from
long-lived progenitor systems. Unfortunately, there are no good {\it H}
luminosities available for large samples of galaxies. 
Since the total luminosity depends on the galaxy distances
the supernova rates depend on $H_0^2$.

Two complementary efforts can be distinguished. One approach is to
collect as many supernovae as possible and then define the galaxy sample
from which they emerged (Tammann et al. 1994). 
This is hampered by the fact that galaxies
without supernova enter the sample only according to some selection
criteria (e.g. contained in a given volume). The other approach is to
only include supernovae which have been detected in a pre-defined galaxy
sample (Cappellaro et al. 1997). This restricts the number of supernovae 
significantly. A new approach, which is tailored for more distant
searches, is to define a galaxy luminosity function and use this
information (as a function of redshift) to determine the rates (Pain et
al. 1996, Reiss 1999). This may be the most efficient way to treat this
problem avoiding massive redshift surveys which cover all galaxies
accessible in the search area and may well require prohibitive
amounts of observing time.

Other problems affecting SN rates are the internal extinction in the
host galaxy. This extinction depends on the local environment and may 
differ considerably for the various supernova types. The extinction of course 
depends on the parent population of the supernovae.
SNe~Ia presumably originate in an older population where extinction
should be less of a problem (but see below
and section \ref{sec:diff} for doubts on the universality of this assumption). 
Unless the age and initial mass function of the supernova parent population 
is the same as that of the dominant stellar population, the assumption 
that the {\it B} luminosity may be a good comparison is questionable. 

The rates of SNe~Ia are very low with about 1 event every 500 to 600 years 
for a galaxy with $10^{10}$L$_{B_{\odot}}$ and a Hubble constant of
65~km~s$^{-1}$~Mpc$^{-1}$ (Cappellaro et al. 1997, Reiss
1999). There is a dependence on galaxy type which shows that SNe~Ia
are observed less often in early type (elliptical and S0) galaxies than
in spirals. This goes against the claim that they emerge from very old
stellar populations. A further puzzling fact questioning the old
paradigm is that there seems to be some preference of SNe~Ia in star
forming galaxies to lie in or near spiral arms (Bartunov et al. 1996) or
at least in their vicinity (McMillan \& Ciardullo 1996) which would
make them of intermediate age ($>$0.5 Gyr).
The rate in the field and in galaxy clusters seems to vary 
very little (Reiss 1999). 

The rates of distant ($z>0.3$) supernovae have been derived only for two very
small samples (Pain et al. 1996, Reiss 1999). 
For the distant supernovae the difficulties in calculating rates
are exacerbated by the small number statistics of spectroscopically
classified objects.
An additional factor for the distant searches are the exact galaxy
luminosities and an approach over some general luminosity functions is
required (Reiss 1999). The restriction to a given filter passband (mostly
the {\it B} band) becomes questionable as there are significant color
changes for distant galaxies.
It is hence not too surprising that the two estimates are not
concordant. Larger supernova samples at high redshift have already been
observed and new rates will become available soon.

\subsubsection{Light curves}

Light curves form one of the main information sources for all supernovae. 
Typically they are observed in the broad-band optical filters following 
the Bessell (1990) system, which combines the older Johnson (Johnson \& 
Harris 1954) and  
Cousins (1980, 1981) filter passbands. The optical {\it UBVRI} bands have been
observed for bright, nearby supernovae. Observations in the near infrared 
{\it JHK} filters have been obtained for a few supernovae only (Elias et al.
1981, 1985, Frogel et al. 1987, Meikle 2000). 

\begin{figure}[t]
\centering
\psfig{file=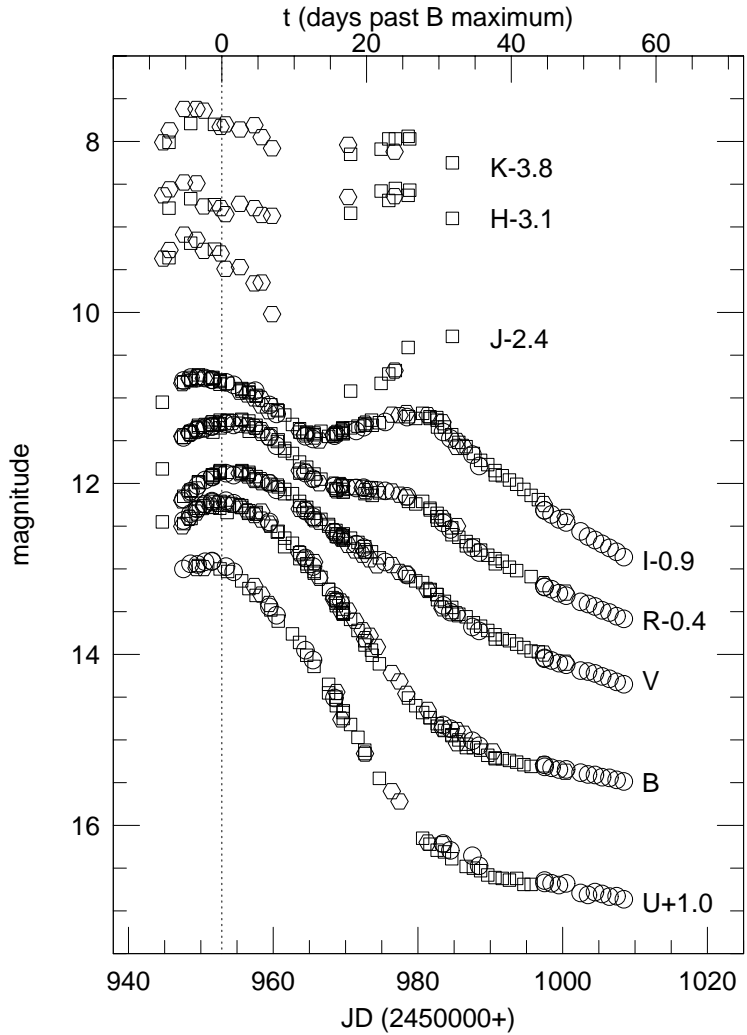,angle=0,%
bbllx=-30mm,bblly=15mm,bburx=275mm,bbury=250mm,height=130mm}
\caption{Optical light and near-infrared light curves of SN~1998bu.
The symbols are for different data sets (circles: Suntzeff et al.
(1999); squares: Jha et al. (1999); hexagons: Hernandez et al.
(2000)). }
\label{fig:98bu-lc}
\end{figure}

Figure~\ref{fig:98bu-lc} displays the
characteristic shape of SNe~Ia in the various filters (Suntzeff et al.
1999, Jha et al. 1999, Hernandez et al. 2000).
Observers usually use the {\it B} maximum as the zero-point for the light curves. 
We will follow this practice here as well.

The light curves have been investigated in detail over the last
decade. After the early assumption of a single time evolution (Minkowski
1964, Leibundgut 1988, Leibundgut et al. 1991a) clear differences emerged 
when new objects were observed in more detail. A striking example of
the differences has been demonstrated by Suntzeff (1996) with the {\it R}
and {\it I} light curves. Earlier indications of deviant objects had been
ignored (Phillips et al. 1987, Frogel et al. 1987, Leibundgut 1988). 

The exact, objective description of optical light curves has become an
industry (Hamuy et al. 1996d, Riess et al. 1996a, Vacca \& Leibundgut
1996, Perlmutter et al. 1997). However, no overall agreement has emerged yet.

\paragraph{Rise Times}
\label{sec:rise}

SNe~Ia rise to maximum very fast. Only in very lucky occasions have early
observations been recorded. One such case has been the occurrence of a
second SN~Ia within 100 days in the same galaxy (SN~1980N and
SN~1981D - Hamuy et al. 1991). The new supernova could be detected on
the deep photographic plates obtained to follow the late light curve
of the first object. Thus the earliest observations were obtained 15.3
days before {\it B} maximum light was reached. Other observations this early
were reported for a small number of SNe~Ia (SN~1971G:
-17 days, Barbon et al. 1973b; SN~1962A: -16 days, Zwicky \& Barbon
1967; SN~1979B: -16 days, Barbon et al. 1982; SN~1999cl: -16 days,
Krisciunas et al. 2000).

Densely sampled supernova searches provide the best chance to obtain 
very early observations. The current Lick Observatory Supernova
Search has already discovered several objects very early (SN~1994ae;
-13 days, Riess et al. 1999a).

A systematic search for early supernova data has been conducted by
Riess et al. (1999b). The earliest reported 
observations are for SN~1990N (-17.9 days) and SN~1998bu (-16.7 days). 
It is thus clear that SNe~Ia rise to {\it B} maximum in more than 18 days.
The rise is very steep with about half a magnitude per day brightness 
increase until about 10 days before maximum (SN~1990N; Riess et al. 1999b).

The pre-maximum light curve is often approximated with a t$^2$ function
(Riess et al. 1999b, Aldering et al. 2000) assuming an expanding fireball
with a very slowly changing temperature. The fit to the data demonstrates
the suitability of such an assumption. Riess et al. find a rise time of
about $-19.5$ days for SNe~Ia. The rise time determined 
by Vacca \& Leibundgut (1996) and Contardo et al. (2000) were based on a
less extended data set and a different functional form. 

\paragraph{Maximum phase}

The maximum phase starts about 5 days before the peak in the {\it B}
filter. At this time a SN~Ia has most likely reached its maximum
brightness in the near-IR filters {\it JHK} (Meikle 2000). We
currently have IR observations for only one SN~Ia at these early
phases, SN~1998bu (Meikle \& Hernandez 1999, Hernandez et al. 2000). 
A dip in the light curve
about 10 days after {\it B} maximum had been observed in {\it JHK}
for other SNe~Ia (Elias et al. 1985, Meikle 2000). SN~1986G possibly had the 
IR maxima observed just a few days before the {\it B} maximum 
(Frogel et al. 1987).

Although there is quite a range 
in relative epochs of maximum in the different filters
it is clear that in most cases SNe~Ia reach maximum earlier
in {\it I} than in {\it B} (Contardo et al. 2000). 
The one object which clearly deviates is SN~1991bg.
It reached {\it I} maximum about 6 days
after the {\it B} maximum (Contardo et al. 2000). 
This is in striking contrast with the other
object reported to be in a similar class, SN~1997cn, which reached
maximum in all filters within a couple of days (Turatto et al. 1998).

The peak phase can be approximated fairly well by Gaussian curves
(Vacca \& Leibundgut 1996, Contardo et al. 2000, Pinto \& Eastman
2000) or second-order polynomials (Hamuy et al. 1996d, Riess et al.
1999a).

The colors evolve very rapidly and non-monotonically around maximum. 
While they appear
fairly constant during the pre-maximum phase, they change from blue
($B-V\approx -0.1$) at 10 days before to red ($B-V\approx 1.1$) 30 days 
after maximum. Other colors evolve similarly, although not as strongly
($V-R$, $R-I$, Ford et al. 1993). A very strong color evolution can be
seen in $J-H$ (from -0.2 to 1.3), while the $H-K$ changes only mildly
(from 0.2 to -0.2; Elias et al. 1985, Meikle 2000), the only color where
SNe~Ia become bluer. In this color the
difference between individual supernovae can be substantial (Meikle
2000). At maximum the typical absorption corrected $B-V$ is about 
$-0.07\pm0.03$. The
$V-I$ color is $-0.32\pm0.04$ with a slight dependence on the
light curve shape (Phillips et al. 1999).

After maximum the supernovae start to fade slowly and go into a
decline at UV and blue ({\it U} and {\it B}) wavelengths. The redder
wavelengths progressively show a decrease of the decline after about
20 days ({\it V}), to a shoulder ({\it R}) and a second maximum ({\it IJHK}). 
The epoch of the second maximum in {\it I} also correlates with other 
parameters, in particular the decline rate and the peak luminosity 
(Suntzeff 1996, Hamuy et al. 1996d, Riess et al. 1996a).

\paragraph{Second maximum}

The characterization of the decline is
not easy and several methods have been proposed. Only the densely
sampled and accurate photometry which became available in the last
decade has allowed us to explore this part of the light curve more
systematically.

A pronounced second maximum has been observed in the {\it I} and redder
light curves (Ford et al. 1993, Suntzeff 1996, Lira et al. 1998, Meikle
2000). This
has been a rather unexpected feature but had been pointed out already
by Elias et al. (1981, 1985). The second maximum has not been characterized 
formally and its interpretation is still unclear (see section
\ref{sec:theo}). The {\it I} light curve
peaks between 21 days (SN~1994D) and 30 days (SN~1994ae) after the {\it B}
maximum. The peaks, however, with quite some spread, are around 29, 25, 
and 21 days past {\it B} maximum for {\it J}, {\it H}, and {\it K}, 
respectively. 
The rise of the
second maximum is very pronounced and amounts from dip to
maximum to about 0.7 mag in {\it J}, 0.6 mag in {\it H}, and 0.4 mag in {\it K}
(Elias et al. 1985, Leibundgut 1988, Meikle 2000). These values are based 
only on very few
objects and any systematic differences could not be described. It is,
however, striking to see how well the templates fit new data like
SN~1998bu (Meikle 2000).

This second peak has been conspicuously absent in the {\it I}
light curves of SN~1991bg (Filippenko et al. 1992b, Turatto et al. 1996) 
and SN~1997cn (Turatto et al. 1998), although a slight change in the
{\it B} and {\it V} light curve decline rates during this phase has 
been reported 
(Leibundgut et al. 1993). 
The IR light curves of SN~1986G (Frogel et al. 1987)
display only a plateau instead of a well formed peak.

\paragraph{Late declines}

After about 50 days the light curves settle onto a steady decline which
is exponential in luminosity. The decline rates are the
same for basically all SNe between 50 and $\sim$120 days (Wells et al. 1994,
Hamuy et al. 1996d, Lira et al. 1998).
The {\it B} light curves decline by about 0.014
mag/day, the {\it V} by 0.028 mag/day, and {\it I} by 0.042 mag/day. 
Exceptions are SN~1986G (Phillips et al. 1987) and SN~1991bg 
(Turatto et al. 1996). They declined faster in {\it B} (0.019 and 
0.020 mag/day, respectively). The decline in {\it V} is identical for all 
SNe~Ia. In {\it I} SN~1991bg declined marginally slower than other 
SNe~Ia (0.040 mag/day). 
The IR light curves have been observed for only a handful of objects
to about 100 days past maximum (Meikle 2000). The decline rate is fairly 
constant for the few objects where it has been observed. It is 0.043 mag/day in
{\it J} and 0.040 mag/day for {\it H} and {\it K} 
(Elias et al. 1985, Leibundgut
1988). These values are almost entirely based on SN~1972E, SN~1980N,
and SN~1981B. Only SN~1980N, SN~1981B, and SN~1981D have been followed
to about 380 days after maximum and show a more or less exponential
decline out to the last observation (Elias \& Frogel 1983). Data in this
range are clearly missing and these epochs will be important to explore
in the future.

\begin{figure}[t]
\centering
\psfig{file=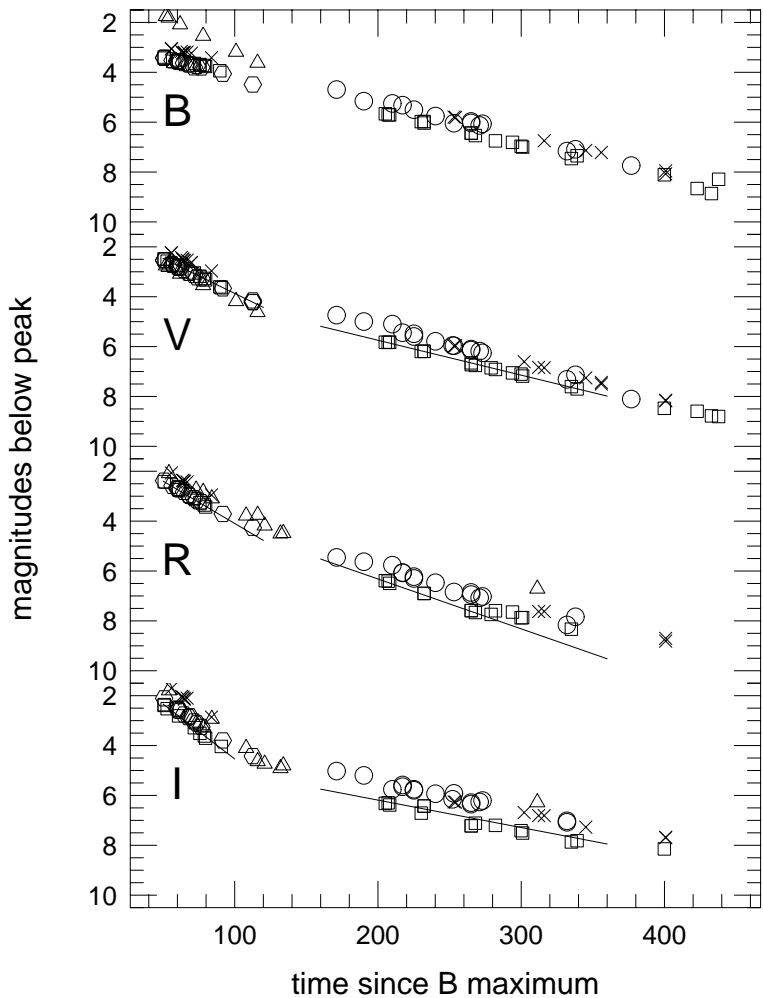,angle=0,%
bbllx=-30mm,bblly=15mm,bburx=275mm,bbury=250mm,height=130mm}
\caption[]{Optical light curves between 100 and 200 days after maximum.
Data of the following supernovae are plotted: SN~1992A (squares),
SN~1994D (hexagons), SN~1991T (crosses), SN~1990N (circles), SN~1986G
(triangles; {\it V} only), and SN~1989B (triangles; {\it R} and {\it I} only).
The lines are fits to the data of SN~1992A.}
\label{fig:late-lc}
\end{figure}

Not many SNe~Ia have been followed much further. At a phase of 150
days past {\it B} maximum a typical supernova is about 5 magnitudes
below its peak brightness and many have disappeared into the glare of
their host galaxy. The few objects which have been observed longer
show a change of slope in the {\it V}, {\it R}, and {\it I} filters 
between 120 and 
140 days (Fig.~\ref{fig:late-lc}; Doggett \& Branch 1985, Lira et al. 1998), 
when the decline
slows to 0.014, 0.015, and 0.011, respectively. The decline rates after
140 days are identical for
SN~1990N, SN~1992A, and SN~1991T (Suntzeff 1996, Lira et al. 1998). 
The {\it B} light curve
maintains its previous slope also at these late phases (Minkowski 1964, Kirshner
\& Oke 1975, Suntzeff 1996, Lira et al. 1998).

A special case is SN~1991T which was observed out to over 1000 days.
A flattening of the {\it B}, {\it V}, and {\it R} light curves after about 600
days was found (Schmidt et al. 1994) and has been observed until 2570 days
after maximum so far (Sparks et al. 1999). The flattening can 
be explained by a light echo produced in a dust layer in front of the 
supernova.

\paragraph{Bolometric light curves}

Given the complex and wavelength-dependent nature of the opacity in
SNe~Ia it is clear that the brightness evolution in individual filter
bands depends on these modulations.
Physically more relevant is the total flux and its change with time. 
Bolometric light curves can provide exactly this. 
Of course, we can not construct fully
bolometric light curves, but only sum over the observed flux. Since this
includes the near-UV, optical and near-infrared wavelengths, such light
curves are often referred to as UVOIR. We will refer to these light
curves as bolometric in the following. Note that we are explicitly
excluding the contributions by $\gamma-$rays. Since most
of the flux emerges in the optical, at least during the first few
weeks, the construction of bolometric light curves is possible
(Suntzeff 1996, Vacca \& Leibundgut 1996, Turatto et al. 1996, 
Contardo et al. 2000).

\begin{figure}[t]
\centering
\psfig{file=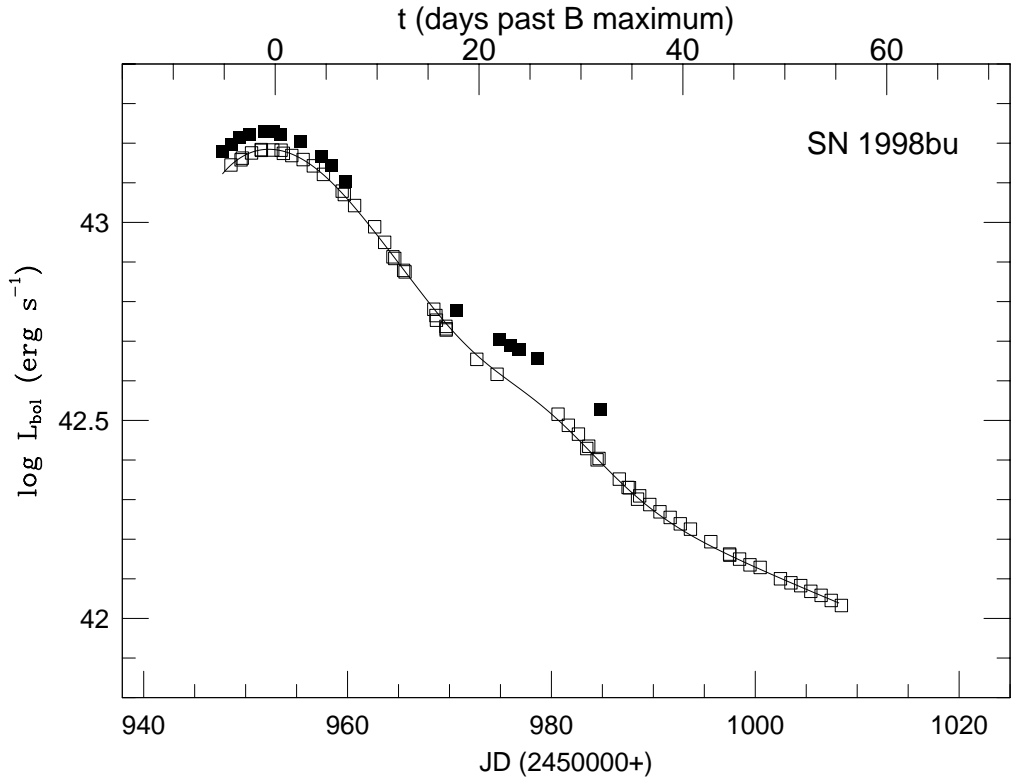,angle=0,%
bbllx=0mm,bblly=15mm,bburx=195mm,bbury=250mm,height=130mm}
\caption[]{Bolometric light curve of SN~1998bu (Contardo 2000). The
open symbols show the {\it UBVRI} integration while the filled squares
display the bolometric light curve including the {\it JHK} bands. The line
is the bolometric light curve derived from fitting the {\it UBVRI} filter
curves individually before integration.}
\label{fig:bol-lc}
\end{figure}

The contribution from the UV is expected to be less than 10\% at maximum
(Suntzeff 1996, Leibundgut 1996) and the IR should also not contribute
significantly.
The published bolometric light curves extend from about 10 days before 
maximum and span a little over 100 days. The
most striking feature is the secondary shoulder which shows up between
20 and 40 days past maximum (Suntzeff 1996, Contardo et al. 2000) in all
SNe~Ia but SN~1991bg. 
We show here the bolometric light curve of SN~1998bu
(Fig.~\ref{fig:bol-lc}; Contardo 2000).
The secondary shoulder is visible about 30 days after maximum. The
contribution of the near-IR passbands {\it JHK} is about 5\% at peak as
predicted and increases through the shoulder as the SN turns redder. 

The peak phase of the bolometric light curve is slightly asymmetric
with the rise from half the peak luminosity being slightly shorter than the
decline to this brightness (Contardo et al. 2000). It takes from 7 to
11 days to double the luminosity before maximum and 10 to 15 days 
to halve it again. The rise to and fall from the maximum is slower for 
more luminous objects. 

The secondary shoulder is visible in many objects, but may vary
considerably in strength and duration.
As with the filter passbands, the shoulder is occurring
later for more slowly declining supernovae (as measured by the decline
to half the luminosity) and hence the more luminous objects (Contardo
2000). 

At late phases SNe~Ia settle onto a decline which is very similar for
all objects, with the exception of SN~1991bg. The decline rate between
50 and 80 days past maximum for the bolometric flux corresponds to
$0.026\pm0.002$ mag/day, while SN~1991bg declined 0.030 mag/day at
this phase.

\subsubsection{Luminosity}

One of the most important ingredients for any analysis of the energetics of 
SNe~Ia is the maximum luminosity. It is also essential for the use of
SNe~Ia as distance indicators and the measurement of the Hubble
constant (Branch \& Tammann 1992, Branch 1998 and references therein).
The best values are currently derived for the few nearby SNe~Ia for which
a distance can be determined by Cepheids. A mean value of
M$_B$=$-19.5\pm0.1$ and M$_V$=$-19.5\pm0.1$ (error of the mean)
for a set of 8 SNe~Ia has been measured (Saha et al. 1999, Gibson et al.
2000). It
has become custom to normalize all SNe~Ia luminosities to a given
decline rate (see section~\ref{sec:diff}). Hence, slightly different
averages can be found for analyses which make differing assumptions on
absorption and perform such a normalization.
A subset of five supernovae treated differently for absorption yields
M$_B$=$-19.7\pm0.1$, M$_V$=$-19.6\pm0.1$ and M$_I$=$-19.3\pm0.1$ 
(Suntzeff et al. 1999),
while another collaboration (Jha et al. 1999) found 
M$_V$=$-19.3\pm0.2$ after the decline rate
correction, which amounts to $\Delta$m$_{corr}$=$-0.26$ globally, 
for four SNe~Ia. The
Suntzeff et al. and Jha et al. absolute magnitudes are hence the same
and differ only marginally from the uncorrected values given in Saha et al.
The discrepancy can be traced to the absorption corrections. Suntzeff
et al. and Jha et al. apply a correction for the host galaxy
absorption which is not done explicitly in Saha et al. 

Apart from the systematic differences on the exact absolute value of
the luminosity it is striking how small the overall scatter of the
measurements is even before the light curve shape
corrections are applied. The total range spans less than 0.5
magnitude in {\it B} and {\it V} (Saha et al. 1999, Gibson et al. 2000). 
It has to be noted that
no Cepheid distance to a truly peculiar object, e.g. SN~1991bg or SN~1991T, 
has been measured so far. The data for NGC~4639 (SN~1991T) have been obtained
and are being analyzed.

The bolometric luminosity of SNe~Ia has been measured for only a handful
of objects. The typical maximum luminosity these objects reach (see
Table~\ref{tab:ni-mass}) is
$10^{43}$~erg~s$^{-1}$ (Contardo et al. 2000). Faint events, like
SN~1991bg, are, however, much less luminous 
($\sim 2\times10^{42}$~erg~s$^{-1}$), 
while the brightest objects reach $> 2\times10^{43}$~erg~s$^{-1}$
(SN~1991T).

\subsubsection{Spectra}

For a recent, very complete, review on the optical spectra of supernovae
of all types see Filippenko (1997a). SNe~Ia are discussed extensively
and readers are referred to this publication for optical spectra (and
references). A large sample of infrared spectra is described in Meikle
et al. (1996) and Bowers et al. (1997).

The evolution of a SN~Ia spectrum is dominated by the changing
influence of various emission and absorption lines. During the early
phases until the late decline in the light curve begins the spectrum
is dominated by P-Cygni lines of intermediate-mass elements. Most
prominent is the Si~{\sc ii} doublet ($\lambda\lambda$6347\AA\ and 6371\AA)
with a prominent absorption of its P-Cygni profile around 6100\AA\
and for a long time the defining feature of SNe~Ia. 
Other prominent lines of SNe~Ia near maximum
light are Ca~{\sc ii} ($\lambda\lambda$3934\AA, 3968\AA, and $\lambda$8579\AA), 
Si~{\sc ii} ($\lambda$3858\AA, $\lambda$4130\AA, $\lambda$5051\AA,  and
$\lambda$5972\AA), 
Mg~{\sc ii} ($\lambda$4481\AA), S~{\sc ii} ($\lambda$5468\AA\ and
$\lambda\lambda$5612\AA, 5654\AA), and O~{\sc i}~($\lambda$7773\AA). 
The spectrum is scattered with
low-ionization Ni, Fe, and Co lines which increase after the peak
(e.g. Jeffery et al. 1992, Mazzali et al. 1993, 1995, 1997). 
Typical velocities observed in the lines are between 10000 and
15000~km~s$^{-1}$.

The spectrum below 3500\AA\ is strongly 
suppressed by lines from iron-peak elements (Harkness 1991, Pauldrach et al.
1996). UV spectra have been obtained for only a few SNe~Ia and only
SN~1990N (Leibundgut et al. 1991b) and SN~1992A (Kirshner et al. 1993)
have regular coverage. All IUE observations are available as a uniform
sample (Cappellaro et al. 1995). 
The features in this part of the spectrum are not due to
regular line formation, but are regions of suppressed line opacity
(Pinto \& Eastman 2000, see also section~\ref{sec:rad}).

The near-IR spectral range is comparatively featureless. Lines of
Si~{\sc ii} ($\lambda1.67\mu$m), Ca~{\sc ii} ($\lambda1.15\mu$m), Mg~{\sc ii}
($\lambda1.05\mu$m) and iron-peak elements 
(between $1.5\mu$m$<$$\lambda$$<$1.7$\mu$m 
and 2.2$\mu$m$<$$\lambda$$<$2.6$\mu$m) are observed (Wheeler et al. 1998). 
A debate on the possible identification of He~{\sc i} ($\lambda1.083\mu$m) 
started with the observations of SN~1994D (Meikle et al. 1996, 
Mazzali \& Lucy 1998). 

There is no appreciable polarization measured in broad-band photometry
and spectra of SNe~Ia (McCall et al. 1984, Spyromilio \& Bailey 1993, 
Wang et al. 1996, 1997). With the exception of 
SN~1996X (Wang et al. 1997) polarized at about 0.2\%, all SNe~Ia have
no detectable polarization in their spectra (Wang et al. 2000).

After the transition from an absorption spectrum, which is superposed on a
pseudo-continuum, to a pure emission spectrum all lines can be
attributed to forbidden Co and Fe transitions (Kirshner \& Oke 1975, Spyromilio
et al. 1992, Kuchner et al. 1994, Bowers et al. 1997, Mazzali et al.
1998, Wheeler et al. 1998). 
The nebular
phase is dominated by the changing strength of these individual line
multiplets. 

\paragraph{Deviations}

Some SNe~Ia have shown significant deviations from the above picture.
Especially SN~1991T (Filippenko et al. 1992a, Phillips et al. 1992,
Jeffery et al. 1992, Mazzali et al. 1995) and SN~1991bg 
(Filippenko et al. 1992b, Leibundgut et al. 1993,
Turatto et al. 1996, Mazzali et al. 1997) have drawn attention to
individual differences among SNe~Ia.

SN~1991T developed the classic Si~{\sc ii} and Ca~{\sc ii} lines very
late and also with diminished strength. Instead, its early spectrum
was dominated by Fe~{\sc iii} lines (Filippenko et al. 1992a, Ruiz-Lapuente
et al. 1992). In the nebular phase SN~1991T was very similar to other
SNe~Ia (Leibundgut et al. 1993) suggesting similar excitation
conditions and densities. The line widths did, however, indicate a
higher expansion velocity (Spyromilio et al. 1992, Mazzali et al. 1998). 

SN~1991bg on the other hand displayed an absorption trough near
$\approx$4000\AA\ which was attributed to Ti~{\sc ii} ($\lambda\lambda$4395\AA,
4444\AA, and 4468\AA) absorption (Filippenko et al. 1992b, Mazzali et al. 1997).
The emergence of this line blend has been explained as a temperature effect
(Nugent et al. 1995).

The stronger lines all show a clear velocity evolution with epoch, which
differs significantly among individual supernovae
(e.g. Branch et al. 1988, Leibundgut et al. 1993, Nugent et al. 1995, 
Patat et al. 1996).
These measurements are
not very reliable as they assume that the expansion velocity can be determined
from the absorption trough of the P-Cygni line. Typical lines analyzed
are the Si~{\sc ii} and the Ca~{\sc ii} doublets. 
High velocity carbon has been inferred from the earliest spectrum of
SN~1990N blended with the Si~{\sc ii} doublet (Fisher et al. 1997). 
The identification is based on the line
profile, with the C~{\sc ii} ($\lambda$6580\AA) line formed in a detached shell.
It is unclear, whether this is a regular feature of other SNe~Ia as well
or was special to SN~1990N.

Line strengths change among individual SNe~Ia as well (Nugent et al.
1995). In particular, a range of Ca~{\sc ii} and Si~{\sc ii} line strengths has
been found. At late phases the line widths also show differences
(Mazzali et al. 1998).

\subsection{Other wavelengths}

There have been attempts to detect nearby SNe~Ia in $\gamma$--rays
with CGRO. These observations would measure the 
$\gamma-$rays from the nuclear decay. The COMPTEL observations of SN~1991T 
have yielded a possible
detection (Morris et al. 1997, Diehl \& Timmes 1998). Deep observations of
SN~1998bu with CGRO have been obtained, but the first reports are
negative. The COMPTEL upper limit clearly excludes the most luminous 
detonation models (Georgii et al. 2000). Also the next
$\gamma$--ray observatory, INTEGRAL will only detect SNe~Ia
closer than about 10 to 15~Mpc depending on the explosion models
(Timmes \& Woosley 1997, H\"oflich et al. 1998a). Prospects for a direct
calibration of he $^{56}$Ni mass hinge on chances for very nearby events.

No X--ray observations have been reported for SNe~Ia. These
supernovae are not expected to emit any significant radiation in this
wavelength regime.

Radio observations of SNe~Ia have been obtained, but no positive
detection has been reported (Weiler et al. 1989, Eck et al. 1995).
A total of 24 SNe~Ia, including all nearby and bright objects,
has been observed at radio wavelengths without a
single detection (Panagia et al. 1999).

\section{Deductions}

This section concentrates on derivatives from the observations and
their possible interpretations.
Recent years have seen several variations emerging from
the formerly very uniform picture. The original assertion that all
SNe~Ia are the same had to be abandoned as better data became
available. In particular, the strong belief in the standard candle
picture, so important for the cosmological applications of SNe~Ia, has
been overthrown by large deviations in light curve shape and spectral
appearance. The monolithic picture of SNe~Ia has been replaced by several
correlations of observable parameters. The full extent of these
correlations has not yet been explored and new ones are still
uncovered.

One of the key questions will be whether SN~1991bg represents an extreme
case in the SN~Ia picture or whether it should be considered separate
and independent of the majority of Ia events. Specifically, it would be
interesting to see whether this object underwent a fundamentally different
explosion (maybe still in the realm of the thermonuclear
explosions) or whether it represents a stripped version of the regular
explosions.

\subsection{Correlations}
\label{sec:diff}

Despite their differences SNe~Ia seem to follow a few invariants in
their appearance. The best-known is the linear decline-rate {\it vs.} 
luminosity
correlation (Phillips 1993). There are now several implementations of this
correction: the template fitting or $\Delta m_{15}$ method (Hamuy et al. 1996b,
Phillips et al. 1999), the multi-light curve shape correction (Riess et
al. 1996a, 1998a), and the stretch factor (Perlmutter et al. 1997). 
Earlier versions of such light curve shape {\it vs.} luminosity relations
had been proposed (Barbon et al. 1973a, Pskovskii 1977, 1984), but could not
be supported by the available data. 

The decline rate correction methods are entirely empirical. They rely on 
the fact that
the fit around the Hubble line in the Hubble diagram improves when they are
applied. Originally based on a set of supernovae with known relative
distances a linear relation for the decline in the {\it B} light curves and
the {\it B}, {\it V}, and {\it I} filter luminosities was determined 
(Phillips 1993). 
It has to be stressed that the relation is normally defined for the 
decline in the {\it B} filter light curve. The coefficients of this relation 
have changed significantly over the last few years as the distances and 
extinction corrections have been refined (Hamuy et al. 1996c,
Phillips et al. 1999; Riess et al. 1996a, 1998a). 
Higher order fits have
now been proposed (Riess et al. 1998a, Phillips et al. 1999, 
Saha et al. 1999).
The stretch factor method describes the light curves by a simple stretch in
time. A basic template light curve is used and then stretched to match
the observations (Perlmutter et al. 1997). This method only works for
the {\it B} and {\it V} light curves through the peak phase, but breaks down
about 4 weeks after peak. It is clear from the {\it R} and {\it I} light curve
shapes that such a stretching procedure can not work linearly for all
filters. Another potential problem with this method is
that it predicts the relative times between filter maxima to stretch
as well, which is not observed. Within these
limitations it has been shown that the stretching provides similar
corrections as the other methods discussed above (Perlmutter et al. 1997;
but see below). 

\begin{figure}[t]
\centering
\psfig{file=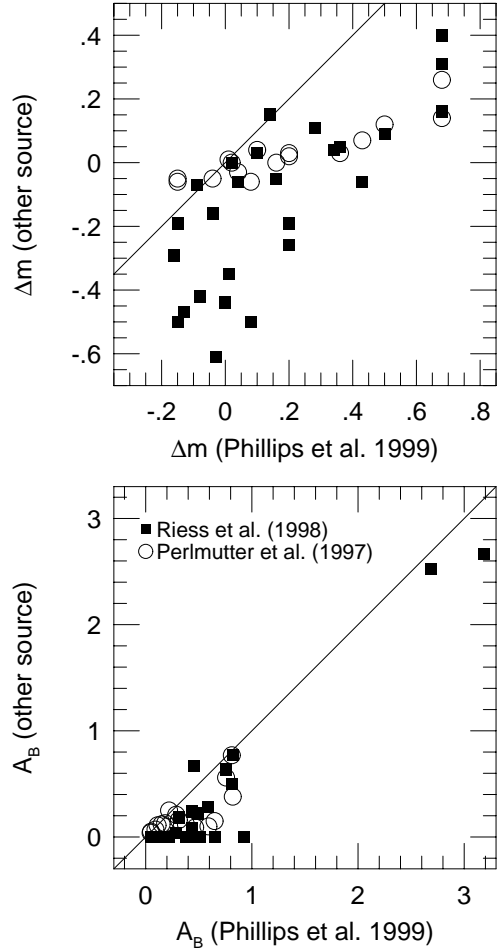,angle=0,%
bbllx=-20mm,bblly=05mm,bburx=275mm,bbury=230mm,height=120mm}
\caption[]{Comparison of the light curve parameters from different
methods. The multi light curve shape (Riess et al. 1998a) and the 
stretch corrections (Perlmutter et al. 1997) are
compared to the template ($\Delta m_{15}$) method for the sample of
nearby supernovae. The top diagram shows
the magnitude corrections based on the light curve shape and the bottom
graph displays the inferred absorption for the different methods.} 
\label{fig:dm15_comp}
\end{figure}

Although some of the methods are equivalent they do 
not reproduce too well. An example is given in Figure~\ref{fig:dm15_comp} (top)
which shows the magnitude corrections determined by the three methods for 
the same supernovae. For the construction of this diagram we have
calculated the magnitude correction given in Phillips et al.
(1999) for the template method. The values for MLCS have been taken 
from Table~10 in Riess et al. 
(1998a) and the stretch correction derived from Table~2 in Perlmutter et al. 
(1999). Note that the assumptions for these magnitude corrections are
not identical for all methods. While Phillips et al. (1999) used {\it
BVI} light curves for their corrections, the ones by Riess et al. (1998a)
and Perlmutter et al. (1999) are based on {\it B} and {\it V} only. 
A significant scatter is noticeable. There is also a significant 
zero-point offset of $0.25\pm0.04$ magnitudes for the MLCS method,
while the stretch yields only a marginal offset of $-0.03\pm0.01$ mag.  
The slopes are further $0.77\pm0.13$ and $0.29\pm0.04$ for MLCS and
stretch, respectively. These are significantly different from the one expected, 
if the methods were identical. The scatter is considerable and also a
concern. A similar conclusion can be drawn from the
comparison of the estimated absorption towards the supernovae
(Fig.~\ref{fig:dm15_comp} bottom). The data are from the same sources
as the magnitude corrections. In
particular, the large spread of absorptions from the template method
compared to both MLCS and stretch determinations (which are based 
on {\it B} and {\it V} only
for these data here) is striking. An overall offset is apparent here as
well.
These are all signatures of subtle, but
significant differences in the treatment of the data. A similar
conclusion has been drawn in connection with data for distant SNe~Ia
(Drell et al. 1999), but attributed to evolution. 

The main differences are
related to the exact correction for dust in the host galaxy. The
determination of the extinction relies on the availability of an
independent absorption indicator. 
In fact, the reddening law in most cases has to be
assumed instead of being derived (for exceptions see Riess et al.
1996b who find a slight deviation from the Galactic reddening law and
Phillips et al. (1999) who don't). 
The best that could be done in the past was to assume that SNe~Ia all
have the same (Branch \& Tammann 1992 and discussion therein)
or a well-defined color (Riess
et al. 1996a) at maximum. With this assumption and the
application of the Galactic reddening law, absorptions could be
measured. Phillips et al. (1999) recently proposed to use the color at
the transition to the nebular phase at an epoch of about 30 days past
peak rather than the color at maximum.
At the transition all supernovae are at about the
same epoch since explosion and the various iron-element emission lines,
which dominate the spectrum, have the same relative strengths. The peak
colors depend on the exact changes of the optical depth in
the ejecta. The color evolution at maximum is also very rapid and small
measurement errors can result in systematically wrong reddening
estimates.

An attempt has been made to combine the known relations into a method
for distance determinations from minimal observations (Riess et al.
1998b). With a single spectrum during the peak phase and photometry
at one epoch in at least two filters the absolute luminosity and the
peak brightness of the event can be determined. The spectrum in this
case provides the information of the phase/age of the supernova and,
through the line strengths, the 'luminosity class,' while the reddening
and brightness are determined from the photometry. The assumption in all
this is, of course, that all SNe~Ia can be described in a single
parameter family and the reddening law is well understood.

An independent fitting method has been applied by Vacca \& Leibundgut
(1996, 1997) and Contardo et al. (2000). Here the data are approximated
with a function which depends on a number of parameters. This fitting
method also reproduces the decline rates found by other methods (Vacca
\& Leibundgut 1997). Its application to larger data sets is still
missing, but would provide a strong independent check on the relations.

There are other parameters which correlate with the peak luminosity of
SNe~Ia. They are the rise time to maximum (Riess et al. 1999b), 
color near maximum light (Riess
et al. 1996a, Tripp 1998, Phillips et al. 1999), line strengths of Ca and 
Si absorption lines (Nugent et al. 1995, Riess et al. 1998b), 
the velocities as measured in Fe lines at late phases (Mazzali et
al. 1998), the host galaxy morphology (Filippenko 1989, Hamuy et al.
1996a, Schmidt et al. 1998), 
and host galaxy colors (Hamuy et al. 1995, Branch et al. 1996). 
There may be indications that the
secondary peak in the {\it I} light curves and also the shoulder in the
bolometric light curves correlate with the absolute luminosity (Hamuy et
al. 1996d, Riess et al. 1999a, Contardo et al. 2000).

\subsection{Energetics}

The observable emission of SNe~Ia is powered completely by the decay of
radioactive $^{56}$Ni and its radioactive daughter nucleus (Colgate \& McKee
1969, Clayton 1974). $^{56}$Ni is synthesized in the explosion and
decays by electron capture with a half-life of 6.1 days to $^{56}$Co.
The cobalt decays through electron capture (81\%) and $\beta^+$ decay (19\%) 
to stable $^{56}$Fe with a half-life of 77 days.
The early phase is dominated by the down-scattering
and the release of photons generated as $\gamma-$rays in the decays
(H\"oflich et al. 1996, Eastman 1997, Pinto \& Eastman 2000). The
dominating opacity comes from the strongly velocity broadened lines and
the incoherent scattering ('line splitting'). 

At late phases the optical radiation is escaping freely and the ejecta
are cooling through emission lines. The $\gamma-$ray escape fraction
increases continually and less and less energy is converted to optical
photons (Leibundgut \& Pinto 1992). After about 150 days
the contribution from positrons becomes significant (Axelrod 1980, Milne
et al. 1999). These positrons
are from a minor channel of the $^{56}$Co decay and annihilate in the
ejecta after losing their kinetic energy in elastic scatterings
(Axelrod 1980, Leibundgut \& Pinto 1992, Ruiz-Lapuente et al. 1995b, 
Milne et al. 1999). 
So far it had been assumed that all the energy from the positron decay
would be deposited in the ejecta, but recent studies indicate that
depending on the magnetic field structure in the ejecta some positrons
could escape and the decay energy is lost for the supernova (Colgate et
al. 1980, Ruiz-Lapuente \& Spruit 1998, Milne et al. 1999).

After several hundred days the emission should change dramatically when
the ejecta has cooled down far enough that the bulk of the cooling
occurs through far-infrared fine structure lines of iron (Fransson et
al. 1996). This has so far not been observed in any SN~Ia as they
are too faint at this point. In some cases the emission becomes
dominated by light echos (e.g. 1991T: Schmidt et al. 1994, Boffi et al. 1999).

\subsection{Progenitors}

All inference of the progenitor systems of supernovae has to come from
the explosion observations themselves. 
The name of the game is to match stellar evolution models with some
parameters indirectly derived from the explosions. Excellent reviews of
this topic are available (Branch et al. 1995, Renzini 1996, Livio 1999,
Nomoto 1999)
and we refer the reader to the references in these compilations.

The observations of SNe~Ia tell us that they emerge from a compact
object as is inferred from the light curves, i.e. the short duration of
the peak phase. The fast decline and the
large leakage of $\gamma-$rays, although only measured indirectly, 
imply a small mass of the ejecta as well.
Note that this statement implicitly made use of the thermonuclear explosion
models (cf. Section \ref{sec:theo}).
The glaring absence of the most abundant elements in the universe,
hydrogen and helium, narrow the selection down to a few highly evolved
objects. Also the appearance in elliptical galaxies with their old
stellar population hints at significant nuclear processing before explosion. 
Indirectly these observational results indicate that SNe~Ia do not
emerge from single star systems. The absence of hydrogen and helium must
mean that the star has removed its envelope. The longevity of the
progenitors, at least for SNe~Ia in elliptical galaxies, and the trigger
of the explosion also point to binary systems.

An important piece of information is the rarefied environment in which SNe~Ia
explode. Radio emission is the telltale signature of such circumstellar
material, but has not been detected in a single SN~Ia. This also sets
limits on mass-loss from companion stars. A potentially powerful
analysis tool would be the detection of narrow H$\alpha$ or He emission from
gas shed by the companion star. The best observations to test for such
emission have been obtained for SN~1994D without a detection (Cumming et
al. 1996). The investigation of local interstellar
environments of SNe~Ia has so far been inconclusive (Van Dyk et al.
1999), but there are at least four cases which could lie close or within
an area of active star formation.

The rareness of SNe~Ia is a further indicator of a special stellar
evolution scenario leading to these explosions. With so few stars
producing SNe~Ia the selection criterion must be rather severe. 

The uniform appearance is often used as an argument for a restrictive
progenitor base, but in the light of the recent observations, in
particular of the faint objects SN~1991bg and SN~1997cn, this should be
investigated again carefully. On the other hand, the strong
correlations hint at fairly unique progenitor systems allowing for some 
variations.

Most discussions on SN~Ia progenitors try to answer two questions. At
what mass does the white dwarf explode (Chandrasekhar or
sub-Chandrasekhar) and what is the donor star. The first question should
be answerable with exact determinations of the Ni masses (see
\S~\ref{sec:ni-mass}) and the combination of this energy source with the
ejecta energy which can be measured from late-phase line widths (e.g.
Mazzali et al. 1998). With the velocities and the column density to
$\gamma-$rays it should be possible to estimate a total ejecta mass.
So far, we have not been able to derive reliable ejecta
masses at all.

The binary companion to the white dwarf could be either another white
dwarf ('double-degenerate'), in which case the two would merge due to 
orbital energy loss by
gravitational radiation, or a regular 'live' star which 
could be either in its giant or main-sequence phase (Renzini 1996). 
In the second case
that star transfers either hydrogen or helium to the white dwarf which
burns the material in a steady phase at its surface. Such systems are
well known as cataclysmic variables and novae (in the case of a
main-sequence companion) or symbiotic stars (where the companion is a
red giant). Supersoft X-ray sources have been identified as white dwarfs
with steady burning hydrogen shells (see Kahabka \& van den Heuvel 1997
for a recent review). Their potential as progenitors of SNe~Ia depends
on the effectiveness with which the white dwarf can increase its mass
towards the Chandrasekhar mass (Hachisu et al. 1999a, b) or whether 
sub-Chandrasekhar explosions are possible (Renzini 1996). Whether SNe~Ia in
elliptical galaxies can be explained by such systems is
controversial. The critical parameter is the mass of the companion star. To
reach ages of $\sim$10$\times$10$^{9}$~years the companion can not be
more than 0.9 to 1.0~M$_{\odot}$ (Hachisu et al. 1999). For compact
binary supersoft sources the companion mass has to be larger than
1.2~M$_{\odot}$ which would make them too short-lived for progenitors in
elliptical galaxies (Kahabka \& van den Heuvel 1997). 

The double degenerate progenitor models have rebounded in recent years
with the detections of several systems with a life time short enough
(Maxted \& Marsh 1999). One system with a total mass near 
the Chandrasekhar limit has been found (Koen et al. 1998). 
Earlier searches for such binaries had been
unsuccessful (e.g. Bragaglia et al. 1990, Renzini 1996). 
There are several questions connected to such progenitor
scenarios as well. Especially the evolutionary path through two
common-envelope phases is still very uncertain. Detailed calculations
on the final merger are also so far inconclusive and in some cases a
collapse to a neutron star is favored (Saio \& Nomoto 1998, Livio 1999). 
Another outcome of
such mergers would be super-Chandrasekhar explosions. It has been
claimed that SN~1991T with its large Ni mass could be explained this way
(Fisher et al. 1999). 

Another possible way to investigate progenitor systems is to determine
the supernova rate as a function of redshift (Ruiz-Lapuente et al. 1995a, 
Madau et al. 1998, Yungelson \& Livio 1998, Ruiz-Lapuente \& Canal 1998,
Dahl\'en \& Fransson 1999). The average age of the
underlying stellar population will influence the number of progenitor
systems available at any given epoch. It is obvious that these rates
also depend on the cosmological model, the star formation history, and 
other observational effects, like dust obscuration (Yungelson \&
Livio 1999) or inhibition of supernova explosions (Kobayashi et al.
1998). There are no indications of changes of
the SN~Ia rate out to redshifts of 0.5 reported (Reiss 1999), but the
data sets are still very limited.

Note, however, that we do not have to peer deep into the universe to
determine differences in SN rates from different progenitor populations.
The age differences can be worked out in the local neighborhood by
comparing the rates in galaxies of different morphological types. Such a
prediction from the evolutionary models would be very helpful to explain the
comparatively large rate in spiral galaxies (Cappellaro et al. 1997).

Of course, any difference between SNe~Ia at large look back times and in
the local neighborhood would indicate differences in the progenitors.
No large differences have been detected so far (cf. Section~\ref{sec:obs}).
Recent discussions on changes in rise times (Riess et al. 1999c,
Aldering et al. 2000) and
possible color evolutions will have to be followed closely in this
respect.

Without clearer indications from observations the progenitor question
will remain unanswered. Searches will have to concentrate on left
overs in any of the progenitor scenarios. In the case of the hydrogen
transfer from the companion a thin layer of hydrogen should still be on
the surface of the explosion or in the wind of the companion. 
It could possibly be observed as a narrow line. The double-degenerate
case would produce a significant mass range and possibly also asymmetric
explosions depending on the accretion of the companion.

\subsection{Nickel masses}
\label{sec:ni-mass}

Once we are in the situation where we can measure the
amount of nickel produced in the explosion, we will probe the explosion
mechanisms more directly. Since most of the white dwarf is burned to the
radioactive $^{56}$Ni and we are observing the subsequent energy
release, we are probing the most sensitive part of the explosion. 

Observationally the measurement of the nickel mass has become possible
recently with the advent of larger telescopes and the development of
better radiation diagnostics. There are two main routes to nickel masses
in SNe~Ia. One is through the observations of the ashes, i.e. the left
over iron from the decays, in the near-infrared and the other by
obtaining the total luminosity at peak and the application of ''Arnett's
law`` (Arnett 1982, Arnett et al. 1985, Branch 1992), which states that 
the energy released on the surface at maximum light is equal to
the energy injected by nuclear decays at the bottom of the ejecta. 
The reason for this is that the atmosphere is turning optically thin 
(Pinto \& Eastman 2000). 
It is possible that Arnett's law is not exact and the ratio of energy
release and input is not exactly unity, e.g. due to asymmetries or
multi-dimensional effects. However, is has to be expected
that all supernovae show a fairly uniform behaviour and the systematic
differences are likely to be small compared with the uncertainties which
arise from extinction and the lack of accurate distances.

The bolometric luminosity has been determined only for very few objects 
(Vacca \& Leibundgut 1996, Turatto et al. 1996, Contardo et al. 2000).   
Since the total luminosity observed at maximum equals the instantaneous 
radioactive decay energy one simply calculates the amount of nickel and its 
daughter product, cobalt, at this moment. The time between explosion and 
maximum is an input parameter in this calculation, although it does not 
introduce severe uncertainties (Contardo et al. 2000).   
The assumed distances and reddening are the critical parameters in all
these analyses. They are needed for the conversion of the observed flux
to absolute units. We have listed in Table~\ref{tab:ni-mass} the nickel
masses as derived by Contardo et al. (2000).

A third possibility is to calculate an exact energy conversion
from the decays and compare this against the observed late light curves.
The latter approach is complicated by the dependence on the exact models
and carries large uncertainties due to partially unknown influences from
the positron channel in the decay (Axelrod 1980, Cappellaro et al. 1998, 
Milne et al. 1999).

Nickel masses have been derived mostly through the near-infrared
observations of [Fe~{\sc ii}] and [Co~{\sc ii}] lines (Spyromilio et al. 1992, 
Bowers et al.
1997). These lines have the advantage that they are largely single
transitions of low excitation stages and not blended, which is
not the case in optical spectra (Axelrod 1980, Kuchner et al. 1994, 
Mazzali et al. 1998). The low ionization is a further advantage as the
atomic data are more reliable than for the higher ionization lines. 
Critical for this method is the accuracy with which the collision
strengths of the lines are known. The uncertainties, especially for the
[Co~{\sc ii}] and [Co~{\sc iii}] lines, are still substantial.

\begin{table}
\caption{Absolute {\it B} magnitudes and bolometric luminosities. 
The nickel mass is derived from the luminosity for 
a rise time of 17 days to the bolometric peak.}
\begin{tabular}{l l r r r r} 
SN & (m-M) & A$_B$ & $\log L_{bol}$ & $M_{\mathrm{Ni}}$~ & $t_{+1/2}$ \\
  &   (mag)  & (mag)  &   (erg s$^{-1}$) & ($M_{\odot}$) &  (days) \\
\hline 
SN 1989B  & 30.22 & 1.55~ & 43.06~~ & 0.57~ & ~12.9~ \\
SN 1991T  & 31.07 & 0.67~ & 43.36~~ & 1.14~ & ~14.2~ \\
SN 1991bg & 31.26 & 0.29~ & 42.32~~ & 0.10~ & ~~8.9~ \\
SN 1992A  & 31.34 & 0.07~ & 42.88~~ & 0.37~ & ~10.6~ \\
SN 1992bc & 34.82 & 0.09~ & 43.22~~ & 0.84~ & ~13.2~ \\
SN 1992bo & 34.63 & 0.11~ & 42.91~~ & 0.41~ & ~~9.9~ \\
SN 1994D  & 30.68 & 0.09~ & 42.91~~ & 0.41~ & ~10.4~ \\
SN 1994ae & 31.86 & 0.63~ & 43.04~~ & 0.55~ & ~12.9~ \\
SN 1995D  & 32.71 & 0.41~ & 43.19~~ & 0.77~ & ~12.9~ \\
SN 1998bu & 30.37 & 1.48~ & 43.18~~ & 0.77~ & ~13.1~ \\
\hline 
\end{tabular}
\label{tab:ni-mass}
\end{table}

The determinations of the nickel masses are fairly consistent between the
different approaches (Contardo et al. 2000).
The major uncertainties are the distances to the supernovae,
which directly influences the luminosities, both for the lines as well
as the bolometric flux.

\section{Theory}
\label{sec:theo}

The detailed theoretical understanding of Type Ia Supernovae is still
limited. Two very complicated physical processes are at work in SNe~Ia
explosions. First there is the explosion mechanism itself, which is
still debated and several possibilities are proposed and then there is
the complicated, highly non-thermal process of the radiation escape which
leads to the observed phenomenon. A recent review of SN~Ia theory is
presented by Hillebrandt \& Niemeyer (2000).

\subsection{Explosion models}
\label{sec:exp}
In general it is agreed that SNe~Ia are the result of thermonuclear
explosions in compact stars. White dwarfs are favored by their intrinsic
instability at the Chandrasekhar mass and the fuel they provide in
carbon and oxygen. All the arguments for this scenario have been already
clearly laid out before 1986 (Woosley \& Weaver 1986 and references
therein). Other fuels could be imagined, but all of them have
some problems. They either do not provide enough (explosive) energy 
(like hydrogen)
or can not synthesize the intermediate-mass elements (like helium, which
detonates). Higher elements are in principle possible, but it is well
known that O-Ne-Mg white dwarfs would rather collapse to a neutron star
than explode because of the large electron capture effects (e.g.
Nomoto \& Kondo 1991, Guti\'errez et al. 1996).
The initiation of the burning in the degenerate star is, however, a puzzle.
For many years it was clear that a detonation (supersonic burning front)
would lead to an
overabundance of iron-group elements and not enough of the
intermediate-mass elements observed in the spectral evolution during the
peak phase. A deflagration (subsonic burning) seemed more appropriate,
but it was not clear how to prevent the explosions to turn into a
detonation. The
phenomenological model W7 (Nomoto et al. 1984, Thielemann et al. 1986) or
similar explosions (Woosley \& Weaver 1986, 1994b) enjoy
a great popularity as the explosive input model for spectral
calculations since they seemed to reproduce the element distribution fairly
accurately (e.g. Harkness 1991, Jeffery et al. 1992, Mazzali et al.
1993, 1995, 1997, Yamaoka et al. 1992, Shigeyama et al. 1992). 
The burning speed in this model has, however, never been understood 
in physical terms. 
Possible alternatives are the pre-expansion of the white dwarf to lift
the degeneracy by a slow deflagration first and have the detonation start 
later (Khokhlov 1991). The critical parameters in these models are the
density at the transition from deflagration to detonation, 
the pre-explosion density, the chemical
composition (mostly C/O ratio), and the deflagration speed at the
beginning of the burning. The transition density has been proposed as
the critical parameter for the nucleosynthesis and hence the amount of Ni 
produced in the explosion.
These delayed-detonation models can reproduce some of the observations
(H\"oflich 1995, H\"oflich \& Khokhlov 1996, H\"oflich et al. 1996). 
However, their
consistency has been questioned recently (Niemeyer 1999, Lisewski et al.
1999a, b). Another
possibility is that the first explosion in the center fizzles and as
the star contracts again, the density and temperatures rise high enough
to re-ignite carbon near the center and lead to the explosion
(Arnett \& Livne 1994a, b, H\"oflich et al. 1995). 
There are hence several
theoretical possibilities to ignite the white dwarf, but it is still not
clear which ones are realized in nature. With the variety of
SN~Ia events observed now, it is possible that SNe~Ia come from
different burning processes. However, the observed correlations must
then be valid across different explosion mechanisms.

Once the explosion has started, the flame has to continue burning enough
material to unbind the star. In many calculations this has not occurred
and the flame has fizzled. Only recently have some three-dimensional 
calculations led to weak explosions (Khokhlov 1995, Niemeyer et al.
1996, Reineke et al. 1999). 

An altogether different explosion mechanism on sub-Chandrasekhar mass white 
dwarfs has been explored (Nomoto 1982, Livne 1990, Livne \& Glasner 1991, 
Woosley \& Weaver 1994a, Livne \& Arnett 1995). 
In this model, the explosion is generated at the surface
of the white dwarf due to a detonation of He at the bottom of the
accretion layer. This model solved the progenitor problem by allowing
explosions well below the Chandrasekhar mass near the peak of the white
dwarf mass distribution. Difficulties here are the initiation of the
explosion and the subsequent ignition of the whole star by a pressure
wave. Many of these calculations are still parametric and the details
have to be worked out (cf. Woosley 1997).

It is customary nowadays to explore several of these explosion models
to explain the observations (e.g. Leibundgut \& Pinto 1992, 
H\"oflich et al. 1995, H\"oflich \& Khokhlov 1996, H\"oflich et al. 1996). 

\subsection{Radiation transport}
\label{sec:rad}
Another complicated process stands between the explosion models and the 
observations. The release of the photons from the explosion is
computationally extremely difficult to follow. The reasons are the
continuous change of the energy deposition and the detailed physics of the
conversion of the $\gamma-$rays injected inside the ejecta from the
radioactive decays to the low-energy photons observed. The opacity
changes due to the thinning of the expanding ejecta for the high-energy
input, but at the same time the high velocities and the abundance of
higher elements with their large number of transitions complicates the
calculations (Harkness 1991, H\"oflich et al. 1993, Eastman 1997, 
Pinto \& Eastman 2000). The
exact treatment is still debated, but it has become increasingly clear
that the old assumption of a thermal input spectrum is not tenable. Even
though SNe~Ia display a nearly thermal 'continuum' during their peak
phase, they are really dominated by the time-dependent photon
distribution. The clearest demonstrations of this fact are the lack of
photons in the {\it J}-band (Spyromilio et al. 1994, Meikle 2000) which is
due to the absence of emission lines in this wavelength region and the
occurrence of the maximum in different optical filters, which is
reversed for most supernovae, i.e. the near-IR filter curves peak before
the optical ones (Contardo et al. 2000, Hernandez et al. 2000). 

Due to the large opacities in the ejecta the photon degradation proceeds
through several channels (e.g. Lucy 1999, Pinto \& Eastman 2000). Since the UV
region is blocked by many velocity-broadened iron-group lines (Harkness
1991, Kirshner et al. 1993), the photons are progressively redshifted
until the optical depth is small enough for them to escape. This occurs
first in the near-IR and hence the peak is reached earlier at these
wavelengths (Meikle 2000, Contardo et al. 2000).
However, only in wavelength regions where plenty of line transitions in
the outer layers are available is there any significant flux. 

Nevertheless, the optical spectrum has been modeled rather successfully
even with thermal input sources 
(Harkness 1991, Jeffery et al. 1992, Mazzali et al. 1993, 1995, 1997, 
Nugent et al. 1997). This is possible since the
outer layers already encounter a pseudo-thermal input spectrum (Pinto
\& Eastman 2000). Detailed treatment of the
NLTE effects has been included by several groups (Baron et al. 1996,
Pauldrach et al. 1996, H\"oflich 1995, H\"oflich et al. 1996, Lucy 1999,
Pinto \& Eastman 2000). 

At late phases the ejecta are optically thin for optical and infrared 
photons and we
see a spectrum dominated by collisionally excited Fe and Co lines 
(Axelrod 1980, Ruiz-Lapuente
\& Lucy 1992, Spyromilio et al. 1992, Kuchner et al. 1995, 
Bowers et al. 1997, Mazzali et al. 1998). At these epochs
it has been assumed that the energy of the positrons in the $^{56}$Co
decay is locally deposited. This has recently been questioned because of the
increased slope of the light curves (Ruiz-Lapuente
\& Spruit 1998, Cappellaro et al. 1998, Milne et al. 1999).
After about 450 days a thermal instability develops in the ejecta which
rapidly cool down from about 3000~K to 300~K. 
Excitation of optical and near-infrared transitions declines
rapidly and the 
cooling continues by fine-structure lines of Fe in the mid- and far-infrared.
This is often referred to as the IR catastrophe. The
predictions are that this would happen after about 500 days (Fransson et
al. 1996) but it has never been observed so far.

It will take a few more years until these problems can be addressed
completely. A closer link between the observations and the models has
been pursued by trying to understand the correlations which have been
observed. The light curve decline has been modeled (H\"oflich et al.
1996) and explained as due to differences in the amount of Ni produced in the
explosion. Also the color dependence could possibly be explained
this way. Other issues like the rise time or the occurrence of the
secondary peak in the near-IR remain, however, open. A possible
interpretation of the light curve stretching during the peak phase and
for the bolometric light curves links the time scales of the Ni decay,
the diffusion time (for a constant opacity) and the age of the
supernova (Arnett 1982, Arnett 1999). 

By comparing the kinetic energy as derived from line widths and the 
measured Ni masses it should be 
possible to derive global parameters of the explosion. First such steps
have been made by Mazzali et al. (1998), Cappellaro et al. (1998), and
Contardo et al. (2000). This alternative route will not replace the
detailed modeling of light curves and spectra, but may provide a more
direct input for the explosion models. 

\section{Discussion}

Although we can foresee a time when there will be more SNe~Ia at
redshifts above 0.3 than nearby ones, we will have to learn from the nearby
samples with their superior data coverage and quality. The distant
supernovae are still observed rather sparsely and lack the wavelength
and spectral coverage we can obtain for local events.

The capabilities for detailed supernova studies have increased
continuously over the past decade and 
detailed spectroscopic and photometric data sets will become available 
at a rapid rate. This will allow
us to address very specific questions and focus on model predictions.
However, simple model predictions have been lacking so far and the
complications in the explosion models and the radiation transport have
proven to be veritable road blocks. 

With the extensive and homogeneous data sets which have become
accessible in the last few years the general discussion of global
parameters of SNe~Ia is possible. The detailed statistics of
luminosity, rise times, decline rates, and spectral line evolutions have
made the systematic investigations of supernova energetics,
explosive nucleosynthesis, and more detailed inferences on progenitors
a possibility. The increased and coordinated access to telescopes of all 
sizes has brought the field forward substantially. With the statistics
on global supernova parameters the tedious comparison with explosion
models has been supplemented.

In the following an attempt is made to assemble the available
information from the observations to point out future directions of SN~Ia
research.

\subsection{Correlations}
\label{sec:disccor}

The differences of the various luminosity corrections and absorption
determinations (see section
\ref{sec:diff}) are very disconcerting. They clearly are not due to
evolution, but will
likely be traced to technicalities of the fits. The most obvious culprit
is the degeneracy of reddening and intrinsic color of SNe~Ia, which has
to be lifted for a reliable measurement. The discrepancies apparent in
Fig.~\ref{fig:dm15_comp} and discussed in section~\ref{sec:diff} are
most likely due to this degeneracy. The
influences on the cosmological conclusions drawn from SNe~Ia will have
to be investigated in much more detail and the discussion has already 
started (e.g. Drell
et al. 1999). The different corrections also play an important role in
the exact determinations of the Hubble Constant (Suntzeff et al. 1999,
Jha et al. 1999, Saha et al. 1999, Gibson et al. 2000) and are
responsible for the remaining discrepancies.

Despite the technicalities of the light curve fitting
the variations among SNe~Ia are real and have to be explained.
It is very interesting to consider the invariants among SNe~Ia. Each
of them tells a different part of the overall story and should help in
piecing it together.

With the brighter SNe seemingly
rising and declining more slowly this implies that the energy release 
is retarded throughout the maximum phase.
The width of the peak phase in the bolometric light curve also
correlates with the peak luminosity (Contardo et al. 2000) and so does
the occurrence of the shoulder (Contardo 2000). The more luminous
objects are so through the whole known evolution, i.e. they are
emitting more energy than the fainter supernovae. It is also
striking to see that the very luminous SNe~Ia show the characteristic
Si~{\sc ii} line appear relatively late (typically only after maximum light). 

The color at maximum is a measure of the opacities in the
ejecta. Since these are dominated mostly by lines and not continuum
processes, the colors are not a direct indicator of the temperature in
the ejecta (Pinto \& Eastman 2000). Interestingly, the $B-V$ color is very
well defined and depends very little on the light curve shape whereas
$V-I$ shows a stronger dependence on the decline parameter (Phillips
et al. 1999). 

The velocities of the ejecta can be measured from the emission lines several
months past the explosion (Mazzali et al. 1998). The fact, that they
correlate with the light curve decline is remarkable. The ejecta
velocities derived from the nebular lines are an indicator of the ratio of
kinetic energy and total mass. Strictly speaking, this only applies
exactly for spherically symmetric explosions, but at the late phases any
asymmetry should be damped out. Since the decline correlates with the
maximum luminosity and this in turn is connected to the Ni mass
synthesized in the explosion (Arnett 1982, Arnett et al. 1985, Branch
1992) we have a direct connection of the explosion strength and a product
of the power source of the supernova emission. More luminous SNe~Ia also 
are more powerful explosions. The ejecta mass of these powerful
explosions must be higher as well, as the slower release of the energy
can only be achieved by a larger column density. This immediately rules
out a single mass progenitor system.

The host galaxy morphology and the galaxy color provide information on
the star formation of the parent population. SNe~Ia in elliptical
galaxies are on average less luminous than their counterparts in late
spirals (Filippenko 1989, Hamuy et al. 1995, Schmidt et al. 1998). 
Also SNe~Ia in bluer galaxies seem to be
brighter (Branch et al. 1996). All of the most luminous objects (like SN~1991T)
occurred in spiral galaxies, and most of them suffer reddening in the
host galaxy and are possibly loosely connected to star forming regions
or spiral arms (Bartunov et al. 1994). It
thus appears as if the more luminous objects are connected to a younger
parent population and the fainter SNe~Ia come from old progenitor
systems. Yet, the correction from the decline rates applies to all
SNe~Ia independent of the host galaxy morphology (Schmidt et al. 1998).
Thus, although the parent population may be different, most likely due
to age differences, most, if not all, SNe~Ia come from the same
type of progenitor system. It has been proposed that the explosion energy
depends on the precursor composition and could be observed from
samples which span sufficiently long look back times, i.e. shorter
progenitor life times (H\"oflich et al. 1998b, Kobayashi et al. 1998,
Umeda et al. 1999). Such experiments will require better and more
extended data than are currently available.

The question of what is a normal SN~Ia has been raised many times in
the last few years (e.g. Branch et al. 1993). Selections based on
color or spectra have been proposed and used. To what extent such
subclassifications describe physical differences is, however, unclear. 

The extreme cases
of SN~1991bg and SN~1997cn can not yet be accommodated within the simple
schemes proposed. Could it be that they emerge from different
mechanisms? The answer is still outstanding and we need more
objects of this sort to investigate. 

\subsection{Nickel masses}

The nickel mass of individual supernovae differs by up to factors of
several (section~\ref{sec:ni-mass}). With such large differences, it is
clear that the explosions are not as uniform as assumed only a decade
ago. It will be an important task for the next years to determine
whether these differences are due to different explosion mechanisms
(e.g. double detonations, deflagrations, etc.) or are variations of a
single mechanism (e.g. the density at transition from deflagration to
detonation (H\"oflich et al. 1996)). 
The distribution of nickel masses may provide a
first indication of what the exact distribution of the explosions is. It
has been claimed that most SNe~Ia emerge from a fairly narrow range of
luminosities (and hence nickel masses), but the numbers are still
small. The distribution of the decline parameters has not yet yielded a
clear picture (e.g. Drell et al. 1999). 

It is of course also possible that we are observing two or more
explosion mechanisms, each with variations on its own. A single explosion
mechanism which produces differences of a factor of 10 as observed in
the nickel masses (Table~\ref{tab:ni-mass}) has not been proposed yet.

\subsection{Future developments}

Three major questions about SNe~Ia will have to be solved:
the influence of reddening, the progenitor systems and the explosion mechanism. 
Observationally,
reddening should be the easiest to either measure or avoid.
Many interesting questions can be addressed with a
statistically significant near-IR sample. The
luminosity corrections are smaller in the {\it I} band (Phillips et
al. 1999), at late times the near-IR is the prefered region for
the determination of the Ni mass (Spyromilio et al. 1992, Bowers et
al. 1997), and the near-IR Hubble diagram
will also provide a reddening free determination of the Hubble constant.
The reddening law in external galaxies will be another topic which
could be addressed by SNe~Ia observed in the optical and the near-IR.

Signatures of SN~Ia progenitors should be discovered soon. 
Either signs of the companion transfer to the white
dwarf are detected or possible progenitor systems can be ruled out on
statistical grounds. Neither has so far been the case. Dedicated
programs for the search of possible progenitor systems are needed.

The constraints on the progenitor models will have to be increased and
improved. The fact that we do not know whether some SNe~Ia come from
sub-Chandrasekhar or Chandrasekhar mass explosions is embarrassing.
In fact, not even the relative ejecta masses are known. If the above
argumentation is correct (section~\ref{sec:disccor}), 
then we have a first sign of explosions with different masses.

The first direct detection of the $\gamma-$rays from the nuclear decay of
$^{56}$Ni and $^{56}$Co would be a major success. Such a detection is
within reach of the current instruments on CGRO or INTEGRAL (H\"oflich
et al. 1998a, Georgii et al. 2000).
The combination of the $\gamma-$ray detection
with the observed (UVOIR) bolometric light curve will be a powerful tool to
measure the escape fraction and hence the energy release in SNe~Ia. 

Statistical studies of the bolometric luminosity of SNe~Ia will further
delineate their true luminosity and hence nickel mass distribution. With
such information it will become possible to decide what the major
stellar evolution
channels for SNe~Ia are. We are entering an interesting phase, where
searches will be volume limited even for faint SNe~Ia and 'fair' samples
can be established. 

The future is bright for SN~Ia research. The extension to high-redshift
searches and the inherent possibility to probe SNe~Ia over significant
look back times offers the opportunity to follow a specific 
tracer of individual stars over a large fraction of the age of the Universe. 
This addition to the current supernova research will tell us about the
history of stars beyond the cosmological implications championed so far.

\section*{Acknowledgments}

Parts of this review have been written during visits at the Astronomical
Institute in Basel and the Stockholm Observatory. I would like to thank
A. G. Tammann and C. Fransson for their hospitality. Many discussions
with M. Phillips, N. Suntzeff, B. Schmidt, R. Kirshner, A. Riess, P.
Meikle, K. Nomoto, W. Hillebrandt, and A. Filippenko are acknowledged. 
Special thanks go to 
G. Contardo for letting me show some of the results of her PhD thesis 
and Jason Spyromilio for continuous critical conversations on supernovae.

\section*{References}
\footnotesize

\begin{list}{}%
{\setlength {\itemindent -10mm} \setlength {\itemsep 0mm} \setlength {\parsep 0mm} \setlength {\topsep 0mm}}

\item Aldering, G., Knop, R., Nugent, P.: 2000, \aj, in press (astro-ph/0001049)
\item Arnett, W. D.: 1982, \apj, 253, 785
\item Arnett, W. D.: 2000, in {\it Supernovae and Gamma-Ray Bursts}, eds. M. Livio, N. Panagia, K. Sahu, Cambridge, Cambridge University Press, in press (astro-ph/9908169)
\item Arnett, W. D., Branch, D., Wheeler, J. C: 1985, Nature, 314, 337
\item Arnett, W. D., Livne, E.: 1994a, \apj, 427, 315
\item Arnett, W. D., Livne, E.: 1994b, \apj, 427, 330
\item Axelrod, T. S.: 1980, in {\it Type I Supernovae}, ed. J. C. Wheeler, (Austin: University of Texas at Austin), 80
\item Barbon, R., Buondi, V., Cappellaro, E., Turatto, M.: 1999, \asas, 139, 531 
\item Barbon, R., Ciatti, F., Rosino, L.: 1973a, \asa, 25, 241
\item Barbon, R., Ciatti, F., Rosino, L.: 1973b, Mem. Soc. Astr. It., 44, 65
\item Barbon, R., Ciatti, F., Rosino, L., Ortolani, S., Rafanelli, P.: 1982, \asa, 116, 43
\item Barbon, R., Benetti, S., Cappellaro, E., Patat, F., Turatto, M.: 1993, Mem. Soc. Astr. It., 64. 1083
\item Bartunov, O. S., Tsvetkov, D. U.: 1997, in {\it Thermonuclear Supernovae},eds. P. Ruiz-Lapuente, R. Canal, J. Isern, Kluwer, Dordrecht, 87
\item Bartunov, O. S., Tsvetkov, D. Y., Filimonova, I. V.: 1994, \pasp, 106, 1276
\item Bessell, M. S.: 1990, \pasp, 102, 1181
\item Bludman, S., Mochkovitch, R., Zinn-Justin, J. (eds.): 1997, {\it Supernovae}, Elsevier, Amsterdam
\item Boffi, F. R., Sparks, W. B., Macchetto, F. D.: 1999, \asas, 138, 253
\item Bowers, E. J. C., Meikle, W. P. S., Geballe, T. R., Walton, N. A., Pinto, P. A., Dhillon, V. S., Howell, S. B., Harrop-Allin, M. K.: 1997, \mnras, 290, 663
\item Bragaglia, A., Greggio, L., Renzini, A., D'Odorico, S.: 1990, \apj, 365, L13
\item Branch, D.: 1992, \apj, 392, 35
\item Branch, D.: 1998, \araa, 36, 17
\item Branch, D., Drucker, W., Jeffery, D. J.: 1988, \apj, 330, L117
\item Branch, D., Tammann, G. A.: 1992, \araa, 30, 359
\item Branch, D., Fisher, A., Nugent, P.: 1993, \aj, 106, 2383
\item Branch, D., Livio, M., Yungelson, L. R., Boffi, F. R., Baron, E.: 1995, \pasp, 107, 1019
\item Branch, D., Romanishin, W., Baron, E.: 1996, \apj, 465, 73 and \apj, 467, 473
\item Burkert, A., Ruiz-Lapuente, P.: 1997, \apj, 480, 297
\item Cappellaro, E., Mazzali, P. A., Benetti, S., Danziger, I. J., Turatto, M., Della Valle, M., Patat, F.: 1998, \asa, 328, 203
\item Cappellaro, E., Turatto, M., Tsvetkov, D. Y., Bartunov, O. S., Pollas, C., Evans., R., Hamuy, M.: 1997, \asa, 322, 383
\item Cappellaro, E., Turatto, M., Fernley, J.: 1995, ESA-SP 1189, ESA, Nordwijk
\item Ciotti, L., D'Ercole, A., Pellegrini, S., Renzini, A.: 1991, \apj, 376, 380
\item Clark, D. H., Stephenson, F.: 1977, The Historical Supernovae, New York, Pergamon Press
\item Clayton, D. D.: 1974, \apj, 188, 155
\item Colgate, S. A., McKee, C.: 1969, \apj, 157, 623
\item Colgate, S. A., Petschek, A. G., Kriese, J. T.: 1980, \apj, 237, L81
\item Contardo, G.: 2000, {\it Monochromatic and Bolometric Light Curves of Type Ia Supernovae}, PhD Thesis, Technical University Munich, Munich
\item Contardo, G., Leibundgut, B., Vacca, W. D.: 2000, \asa, submitted
\item Cousins, A. W. J.: 1980, S. Afr. Astron. Obs. Cir. 1, 116
\item Cousins, A. W. J.: 1981, S. Afr. Astron. Obs. Cir. 6, 4
\item Cumming, R. J., Lundqvist, P., Smith, L. J., Pettini, M., King, D. L.: \mnras, 283, 1355
\item Dahl\'en, T., Fransson, C.: 1999, \asa, 350, 349
\item Diehl, R., Timmes, F. X.: 1998, \pasp, 110, 637
\item Doggett, J. B., Branch, D.: 1985, \aj, 90, 2303
\item Drell, P. S., Loredo, T. J., Wasserman, I.: 2000, \apj, in press (astro-ph/9905027)
\item Eastman, R. G.: 1997, in {\it Thermonuclear Supernovae}, eds. P. Ruiz-Lapuente, R. Canal, J. Isern, Kluwer, Dordrecht, 571
\item Eck, C. R., Cowan, J. J., Roberts, D. A., Boffi, F. R., Branch, D.: 1995, \apj, 451, L53
\item Elias, J. H., Frogel, J. A., Hackwell, J. A., Persson, S. E.: 1981, \apj, 251, L13
\item Elias, J. H., Frogel, J. A.: 1983, \apj, 268, 718
\item Elias, J. H., Matthews, K., Neugebauer, G., Persson, S. E.: 1985, \apj, 296, 379
\item Ferrara, A., Tolstoy, E.: 2000, \mnras, in press (astro-ph/9905280)
\item Filippenko, A. V.: 1989, \pasp, 101, 588
\item Filippenko, A. V.: 1997a, \araa, 35, 309
\item Filippenko, A. V.: 1997b, in {\it Thermonuclear Supernovae}, eds. P. Ruiz-Lapuente, R. Canal, J. Isern, Kluwer, Dordrecht, 1 
\item Filippenko, A. V., et al.: 1992a, \apj, 384, L15
\item Filippenko, A. V., et al.: 1992b, \aj, 104, 1543
\item Fisher, A., Branch, D., Nugent, P., Baron, E.: 1997, \apj, 481, L89
\item Fisher, A., Branch, D., Hatano, K., Baron, E.: 1999, \mnras, 304, 679
\item Ford, C. H., Herbst, W., Richmond, M. W., Baker, M. L., Filippenko, A. V., Treffers, R. R., Paik, Y., Benson, P. J.: 1993, \aj, 106, 1101
\item Fransson, C., Houck, J., Kozma, C.: 1996 in {\it Supernovae and Supernova Remnants}, eds. McCray R. and Wang Z. Cambridge University Press, Cambridge, p. 211
\item Frogel, J. A., Gregory, B., Kawara, K., Laney, D., Phillips, M. M., Terndrup, D., Vrba, F., Whitford, A. E.: 1987, \apj, 315, L129
\item Georgii, R., et al.: 2000, in {\it AIP Proceedings of the 5th Compton Symposium}, in press
\item Germany, L., Reiss, D. J., Sadler, E. M., Schmidt, B. P., Stubbs, C. W.: 1999, \apj, in press (astro-ph/9906096)
\item Gibson, B. K., et al.: 2000, \apj, in press (astro-ph/9908149)
\item Guti\'errez, J., Garc\'ia-Berro, E., Iben, I., Isern, J., Labay, J., Canal, R.: 459, 701
\item Hachisu, I., Kato, M., Nomoto, K.: 1999a, \apj, 522, 487
\item Hachisu, I., Kato, M., Nomoto, K., Umeda, H.: 1999b, \apj, 519, 314
\item Hamuy, M., Phillips, M. M., Maza, J., Wischnjewsky, M., Uomoto, A., Landolt, A. U., Khatwani, R.: 1991, \aj, 102, 208
\item Hamuy, M., Phillips, M. M., Maza, J., Suntzeff, N. B., Schommer, R. A., Avil\'es, R.: 1995, \aj, 109, 1
\item Hamuy, M., Phillips, M. M., Schommer, R. A., Suntzeff, N. B., Maza, J., Avil\'es, R.: 1996a, \aj, 112, 2391
\item Hamuy, M., Phillips, M. M., Suntzeff, N. B., Schommer, R. A., Maza, J., Avil\'es, R.: 1996b, \aj, 112, 2398
\item Hamuy, M., et al.: 1996c, \aj, 112, 2408
\item Hamuy, M., Phillips, M. M., Suntzeff, N. B., Schommer, R. A., Maza, J., Smith, R. C., Lira, P., Avil\'es, R.: 1996d, \aj, 112, 2438
\item Harkness, R. P.: 1991, in {\it Supernovae}, ed. S. Woosley, Springer, Heidelberg, 454
\item Harkness, R. P., Wheeler, J. C.: 1990, in {\it Supernovae}, ed. A. G. Petschek, Springer, New York, 1
\item Hernandez, M., et al.: 2000, \mnras, in preparation
\item Hillebrandt, W., Niemeyer, J. C.: 2000, \araa, 38, in press
\item H\"oflich, P.: 1995, \apj, 443, 89
\item H\"oflich, P., M\"uller, E., Khokhlov, A. M.: 1993, \asa, 268, 570
\item H\"oflich, P., Khokhlov, A. M., Wheeler, J. C.: 1995, \apj, 444, 831
\item H\"oflich, P., Khokhlov, A. M.: 1996, \apj, 457, 500
\item H\"oflich, P., Khokhlov, A. M., Wheeler, J. C., Phillips, M. M., Suntzeff, N. B., Hamuy, M.: 1996, \apj, 472, L81
\item H\"oflich, P., Wheeler, J. C., Khokhlov, A.: 1998a, \apj, 492, 228
\item H\"oflich, P., Wheeler, J. C., Thielemann, F.-K.: 1998b, \apj, 495, 617
\item Iben, I., Tutukov, A. V.: 1994, \apj, 431, 264
\item Iben, I., Tutukov, A. V.: 1999, \apj, 511, 324
\item Jeffery, D. J., Leibundgut, B., Kirshner, R. P., Benetti, S., Branch, D., Sonneborn, G.: 1992, \apj, 397, 304
\item Jha, S., et al.: 1999, \apjs, 125, 73
\item Johnson, H. L, Harris, D. L.: 1954, \apj, 120, 196
\item Kahabka, P., van den Heuvel, E. P. J.: 1997, \araa, 35, 69
\item Kamionkowski, M., Kosowski, A.: 1999, {\it Ann. Rev. Nucl. Part. Sci}, 49, 77
\item Khokhlov, A. M.: 1991, \asa, 245, 114
\item Khokhlov, A. M.: 1995, \asa, 245, 114
\item Kirshner, R. P., Oke, J. B.: 1975, \apj, 200, 574
\item Kirshner, R. P., et al.: 1993, \apj, 415, 589
\item Kobayashi, C., Tsujimoto, T., Nomoto, K, Hachisu, I., Kato, M.: 1998, \apj, 503, L155
\item Koen, C., Orosz, J. A., Wade, R. A.: 1998, \mnras, 300, 695
\item Krisciunas, K., Diercks, A., Hastings, N. C., Loomis, K., Magnier, E., McMillan, R., Riess, A. G., Stubbs, C.: 2000, \apj, submitted (astro-ph/9912219)
\item Kuchner, M. J., Kirshner, R. P., Pinto, P. A., Leibundgut, B.: 1994, \apj, 426, L89
\item Leibundgut, B.: 1988, {\it Light curves of Supernovae Type I}, PhD Thesis, University of Basel, Basel
\item Leibundgut, B.: 1996, in {\it Supernovae and Supernova Remnants}, eds. R. McCray, Z. Wang, Cambridge: Cambridge University Press, 11
\item Leibundgut, B., Kirshner, R. P., Filippenko, A. V., Shields, J. C., Foltz, C. B., Phillips, M. M., Sonneborn, G.: 1991b, \apj, 371, L23
\item Leibundgut, B., Tammann, G. A., Cadonau, R., Cerrito, D.: 1991a, \asas, 89, 537
\item Leibundgut, B., et al.: 1993, \aj, 105, 301
\item Leibundgut, B., Pinto, P. A.: 1992, \apj, 401, 49
\item Li, W. D., et al.: 1999, \aj, 117, 2709
\item Lira, P., et al.: 1998, \aj, 115, 234
\item Lisewski, A. M, Hillebrandt, W., Woosley, S. E., Niemeyer, J. C., Kerstein, A. R.: 2000, \apj, in press (astro-ph/9909508)
\item Lisewski, A. M., Hillebrandt, W., Woosley, S. E.: 2000, \apj, in press (astro-ph/9910056)
\item Livio, M., Panagia, N., Sahu, K. (ed.): 2000, {\it Supernovae and
Gamma-Ray Bursts}, Cambridge, Cambridge University Press
\item Livio, M.: 1999, in {\it Type Ia Supernovae: Theory and Cosmology}, eds.
J. C. Niemeyer and J. W. Truran, Cambridge, Cambridge University Press, in press (astro-ph/9903264)
\item Livne, E.: 1990, \apj, 354, L53
\item Livne, E., Glasner, A. S.: 1991, \apj, 370, 272
\item Livne, E., Arnett, W. D.: 1995, \apj, 452, 62
\item Lucy, L. B.: 1999, \asa, 345, 211
\item Madau, P., Della Valle, M., Panagia, N.: 1998, \mnras, 297, L17
\item Maxted, P. F. L., Marsh, T. R.: 1999, \mnras, 307, 122
\item Mazzali, P. A., Lucy, L. B., Danziger, I. J., Gouiffes, C., Cappellaro, E., Turatto, M.: 1993, \asa, 269, 423
\item Mazzali, P. A., Cappellaro, E., Danziger, I. J., Turatto, M., Benetti, S.: 1998, \apj, 499, L49
\item Mazzali, P. A., Chugai, N., Turatto, M., Lucy, L. B., Danziger, I. J., Cappellaro, E., Della Valle, M., Benetti, S.: 1997, \mnras, 284, 151
\item Mazzali, P. A., Danziger, I. J., Turatto, M.: 1995, \asa, 297, 509
\item Mazzali, P. A., Lucy, L. B.: 1998, \mnras, 295, 428
\item McCall, M. L., Reid, N., Bessell, M. S., Wickramasinghe, D.: 1984, \mnras, 210, 839
\item McCray, R., Wang, Z. (eds.): 1996, {\it Supernovae and Supernova Remnants}, Cambridge: Cambridge University Press
\item McMillan, R. J., Ciardullo, R.: 1996, \apj, 473, 707
\item Meikle, P., Hernandez, M.: 1999, in {\it Future Directions of Supernova Research}, eds. S. Cassisi and P. Mazzali, Mem. Soc. Astr. It., in press (astro-ph/9902056)
\item Meikle, W. P. S., et al.: 1996, \mnras, 281, 263
\item Meikle, W. P. S.: 2000, \mnras, in press (astro-ph/9912123)
\item Milne, P. A., The, L.-S., Leising, M.: 1999, \apjs, 124, 503
\item Minkowski, R.: 1941, \pasp, 53, 224
\item Minkowski, R.: 1964, \araa, 2, 247
\item Morris, D. J., et al.: 1997, in {\it 4th Compton Symposium on Gamma-Ray Astronomy and Astrophyics}, eds. C. Dermer, et al. AIP 410, New York, 1084
\item Mould, J. R., et al.: 2000, \apj, 528, 655
\item Murdin, P., Murdin, L.: {\it Supernovae}, Cambridge University Press, Cambridge
\item Niemeyer, J. C.: 1999, \apj, 523, L57
\item Niemeyer, J. C., Hillebrandt, W., Wooseley, S. E.: 1996, \apj, 471, 903
\item Niemeyer, J. C., Truran, J. W. (eds.): 1999, Type Ia Supernovae: Theory and Cosmology, Cambridge, Cambridge University Press
\item Nomoto, K.: 1982, \apj, 257, 780
\item Nomoto, K.: 2000, in {\it Type Ia Supernovae: Theory and
Cosmology}, eds. J. C. Niemeyer and J. W. Truran, Cambridge, Cambridge University Press, in press (astro-ph/9907386)
\item Nomoto, K., Kondo, Y.: 1991, \apj, 367, L19
\item Nomoto, K., Thielemann, F.-L., Yokoi, K.: 1984, \apj, 286, 644
\item Nugent, P., Phillips, M. M., Baron, E., Branch, D., Hauschildt, P.: 1995, \apj, 455, L147
\item Nugent, P., Baron, E., Branch, D., Fisher, A., Hauschildt, P. H.: 1997, \apj, 485, 812
\item Pain, R., et al.: 1996, \apj, 473, 356
\item Panagia, N., Weiler, K. W., Lacey, C., Montes, M., Sramek, R. A., Van Dyk, S. D.: in {\it Future Directions of Supernova Research}, eds. S. Cassisi and P. Mazzali, Mem. Soc. Astr. It., in press (STScI Preprint 1359)
\item Patat, F., Benetti, S., Cappellaro, E., Danziger, I. J., Della Valle, M., Mazzali, P. A., Turatto, M.: 1996, \mnras, 278, 111
\item Pauldrach, A. W. A., Duschinger, M., Mazzali, P. A., Puls, J., Lennon, M., Miller, D. L.: 1996, \asa, 312, 525
\item Perlmutter, S., et al.: 1997, \apj, 483, 565
\item Perlmutter, S., et al.: 1999, \apj, 517, 565
\item Petschek, A. G. (ed.): 1990, {\it Supernovae}, Springer, Heidelberg
\item Phillips, M. M.: 1993, \apj, 413, L105
\item Phillips, M. M., et al.: 1987, \pasp, 99, 592
\item Phillips, M. M., Lira, P., Suntzeff, N. B., Schommer, R. A., Hamuy, M., Maza, J.: 1999, \aj, 118, 1766
\item Phillips, M. M., Wells, L. A., Suntzeff, N. B., Hamuy, M., Leibundgut, B., Kirshner, R. P., \& Foltz, C. B.: 1992, \aj, 103, 1632
\item Pinto, P. A., Eastman, R. G.: 2000, \apj, in press (astro-ph/9611195)
\item Pskovskii, Y. P.: 1977, Sov. Astron., 21, 675
\item Pskovskii, Y. P.: 1984, Sov. Astron., 28, 658
\item Reineke, M., Hillebrandt, W., Niemeyer, J. C.: 1999, \asa, 347, 747
\item Reiss, D. J.: 1999, {\it The Rate of Supernovae in the Nearby and Distant Universe}, PhD Thesis, University of Washington
\item Reiss, D. J. Germany, L. M., Schmidt, B. P., Stubbs, C. W.: 1998, \aj, 115, 26
\item Renzini, A.: 1996, in {\it Supernovae and Supernova Remnants}, eds. R. McCray and Z. Wang, Cambridge: Cambridge University Press, 77
\item Renzini, A.: 1999, in {\it Chemical Evolution from Zero to High Redshift}, eds. J. R. Walsh, M. R. Rosa, Heidelberg: Springer Verlag, 185
\item Richmond, M. W., Treffers, R. R., Filippenko, A. V.: 1993, \pasp, 105, 1164
\item Riess, A. G., Press, W. H., Kirshner, R. P.: 1996a, \apj, 473, 88
\item Riess, A. G., Press, W. H., Kirshner, R. P.: 1996b, \apj, 473, 588
\item Riess, A. G., et al.: 1998a, \aj, 116, 1009
\item Riess, A. G., Nugent, P., Filippenko, A. V., Kirshner, R. P., Perlmutter, S.: 1998b, \apj, 504, 935
\item Riess, A. G., et al.: 1999a, \aj, 117, 707
\item Riess, A. G., et al.: 1999b, \aj, 118, 2675
\item Riess, A. G., Filippenko, A. V., Li, W., Schmidt, B. P: 1999c, \aj, 118, 2668
\item Ruiz-Lapuente, P., Canal, R., Isern, J. (eds.): 1997, Thermonuclear Supernovae, Kluwer, Dordrecht
\item Ruiz-Lapuente, P., Lucy, L.: 1992, \apj, 400, 127
\item Ruiz-Lapuente, P., Burkert, A., R. Canal: 1995a, \apj, 447, L69
\item Ruiz-Lapuente, P., Canal, R.: 1998, \apj, 497, L57
\item Ruiz-Lapuente, P., Spruit, H. C.: 1998, \apj, 500, 360
\item Ruiz-Lapuente, P., Cappellaro, E., Turatto, M., Gouiffes, C., Danziger, I. J., Della Valle, M., Lucy, L. B.: 1992, \apj, 387, L33
\item Ruiz-Lapuente, P., Kirshner, R. P., Phillips, M. M., Challis, P. M., Schmidt, B. P., Filippenko, A. V., Wheeler, J. C.: 1995b, \apj, 439, 60
\item Saffer, R. A., Livio, M., Yungelson, L. R.: 1998, \apj, 502, 394
\item Saha, A, Sandage, A., Tammann, G. A., Labhardt, L., Macchetto, F. D., Panagia, N.: 1999, \apj, 522, 802
\item Saio, H., Nomoto, K.: 1998, \apj, 500, 388
\item Schmidt, B. P., Kirshner, R. P., Leibundgut, B., Wells, L. A., Porter, A. C., Ruiz-Lapuente, P., Challis, P., Filippenko, A. V.: 1994, \apj, 434, L19
\item Schmidt, B. P., et al.: 1998, \apj, 507, 46
\item Shigeyama, T., Suzuki, T., Kumagai, S., Nomoto, K., Saio, H., Yamaoka, H.: 1994, \apj, 420, 341
\item Sparks, W. B., Maccetto, F. D., Panagia, N., Boffi, F. R., Branch, D., Hazen, M. L., Della Valle, M.: 1999, \apj, 523, 585
\item Spyromilio, J., Bailey, J.: 1993, \pasa, 10, 263
\item Spyromilio, J., Meikle, W. P. S., Allen, D. A., Graham, J. R.: 1992, \mnras, 258, 53
\item Spyromilio, J., Pinto, P. A., Eastman, R. G.: 1994, \mnras, 266, L17
\item Strom, R. G.: 1995, in {\it The Lives of Neutron Stars}, eds. M. A.
Alpar, \"U. Kiziloglu, and J. van Paradijs, Kluwer, Dordrecht, 23
\item Suntzeff, N. B.: 1996, in {\it Supernovae and Supernova Remnants}, eds. R. McCray and Z. Wang, Cambridge University Press, Cambridge, 41
\item Suntzeff, N. B., et al.: 1999, \aj, 117, 1175
\item Suntzeff, N. B.: 1997, private communication
\item Umeda, H., Nomoto, K., Kobayashi, C., Hachisu, I, Kato, M.: 1999, \apj, 522, L43
\item Tammann, G. A.: 1994, in {it Supernovae}, eds. S. Bludman, R. Mochkovitch, and J. Zinn-Justin, Elsevier, Amsterdam, 1
\item Tammann, G. A., L\"offler, W., Schr\"oder, A.: 1994, \apjs, 92, 487
\item Thielemann, F.-K., Nomoto, K., Yokoi, K.: 1986, \asa, 158, 17
\item Timmes, F. X., Woosley, S. E.: 1997, 489. 160
\item Trimble, V.: 1982, Rev. Mod. Phys., 54, 1183
\item Trimble, V.: 1983, Rev. Mod. Phys., 55, 511
\item Tripp, R.: 1998, \asa, 331, 815
\item Turatto, M., Benetti, S., Cappellaro, E., Danziger, I. J., Della Valle, M., Mazzali, P. A., Patat, F.: 1996, \mnras, 283, 1
\item Turatto, M., Piemonte, A., Benetti, S., Cappellaro, E., Mazzali, P. A., Danziger, I. J., Patat, F.: 1998, \aj, 116, 2431
\item Turatto, M., et al.: 1999, in {\it Future Directions of Supernova Research}, eds. S. Cassisi and P. Mazzali, Mem. Soc. Astr. It., in press
\item van den Bergh, S.: 1990, \pasp, 102, 1318
\item van den Bergh, S., Tammann, G. A.: 1991, \araa, 29, 363
\item Van Dyk, S. D., Peng, Y. C., Barth, A. J., Filippenko, A. V.: 1999, \aj, 118, 2331
\item Vacca, W. D., Leibundgut, B.: 1996, \apj, 471, L37
\item Vacca, W. D., Leibundgut, B.: 1997, in {\it Thermonuclear Supernovae}, eds. P. Ruiz-Lapuente, R. Canal, J. Isern, Kluwer, Dordrecht, 65
\item Wang, L., Howell, A., H\"oflich, P., Wheeler, J. C.: 2000, \apj, in press (astro-ph/9912033)
\item Wang, L. Wheeler, J. C., H\"oflich, P.: 1997, \apj, 476, L27
\item Wang, L., Wheeler, J. C., Li, Z., Clocchiatti, A.: 1996, \apj, 467, 435
\item Weiler, K. W., Panagia, N., Sramek, R. A., van der Hulst, J. M., Roberts, M. S., Nguyen, L.: 1989, \apj, 336, 421
\item Wells, L., et al.: 1994, \aj, 108, 2233
\item Wheeler, J. C., Harkness, R. P., Khokhlov, A. M., H\"oflich, P.: 1995, {\it Phys. Reps.}, 256, 211
\item Wheeler, J. C., H\"oflich, P., Harkness, R. P., Spyromilio, J.: 1998, \apj, 496, 908
\item Wheeler, J. C., Piran, T., Weinberg, S. (eds.): 1990, {\it Supernovae}, Singapore, World Scientific Publishing
\item Woosley, S. E. (ed.): 1991, {\it Supernovae}, Springer, Heidelberg
\item Woosley, S. E.: 1997, in {\it Thermonuclear Supernovae}, eds. P. Ruiz-Lapuente, R. Canal, J. Isern, Kluwer, Dordrecht, 313
\item Woosley, S. E., Weaver, T. A.: 1986, \araa, 24, 205
\item Woosley, S. E., Weaver, T. A.: 1994a, \apj, 423, 371
\item Woosley, S. E., Weaver, T. A.: 1994b, in {\it Supernovae}, eds. S. Bludman, R. Mochkovitch, and J. Zinn-Justin, Elsevier, Amsterdam, 63
\item Wyse, R. F. G., Silk, J.: 1985, \apj, 296, L1
\item Yamaoka, H., Nomoto, K., Shigeyama, T., Thielemann, F.-K.: 1992, \apj, 393, 55
\item Yungelson, L., Livio, M.: 1998, \apj, 497, 168
\item Zwicky, F.: 1965, in {\it Stellar Structure}, eds. L. H. Aller \& D. B.
McLaughlin, University of Chicago Press, Chicago, 367
\item Zwicky, F., Barbon, R.: 1967, \aj, 72, 1366

\end{list}

\end{document}